\documentclass[useAMS,usenatbib]{mn2e}  
\usepackage{graphicx,amssymb,afterpage}
\newcommand \hii{H\,{\sc ii}}
\begin{document}
   \title{Modelling the Galactic distribution of free electrons}

   \author[D.H.F.M. Schnitzeler]
          {D.H.F.M. Schnitzeler$^{1,2}$\thanks{E-mail: Schnitzeler@mpifr-bonn.mpg.de}\\
          $^1$ Australia Telescope National Facility, CSIRO Astronomy and Space Science, Marsfield, NSW 2122, Australia\\
          $^2$ Max Planck Institut f\"ur Radioastronomie, 53121 Bonn, Germany
}
   \date{}

   \pagerange{}\pubyear{}

   \maketitle

 
\begin{abstract}
An accurate picture of how free electrons are distributed throughout the Milky Way leads to more reliable distances for pulsars, and more accurate maps of the magnetic field distribution in the Milky Way. In this paper we test 8 models of the free electron distribution in the Milky Way that have been published previously, and we introduce 4 additional models that explore the parameter space of possible models further. These new models consist of a simple exponential thick disk model, and updated versions of the models by Taylor \& Cordes and Cordes \& Lazio with more extended thick disks. The final model we introduce uses the observed H$\alpha$ intensity as a proxy for the total electron column density, also known as the dispersion measure (DM). Since accurate maps of H$\alpha$ intensity are now available, this final model can in theory outperform the other models. We use the latest available data sets of pulsars with accurate distances (through parallax measurements or association with globular clusters) to optimise the parameters in these models. In the process of fitting a new scale height for the thick disk in the model by Cordes \& Lazio we discuss why this thick disk cannot be replaced by the thick disk that Gaensler et al. advocated in a recent paper. In the second part of our paper we test how well the different models can predict the DMs of these pulsars at known distances. We base our test on the ratios between the modelled and observed DMs, rather than on absolute deviations, and we identify systematic deviations between the modelled and observed DMs for the different models. For almost all models the ratio between the predicted and the observed DM cannot be described very well by a Gaussian distribution. We therefore calculate the deviations N between the modelled and observed DMs instead, and compare the cumulative distributions of N for the different models. Almost all models perform well, in that they predict DMs within a factor of 1.5--2 of the observed DMs for about 75\% of the lines of sight. This is somewhat surprising since the models we tested range from very simple models that only contain a single exponential thick disk to very complex models like the model by Cordes \& Lazio. We show that the model by Taylor \& Cordes that we updated with a more extended thick disk consistently performs better than the other models we tested. Finally, we analyse which sightlines have DMs that prove difficult to predict by most models, which indicates the presence of local features in the ISM between us and the pulsar.
\end{abstract}


%

\begin{figure*}
\resizebox{\hsize}{!}{\includegraphics{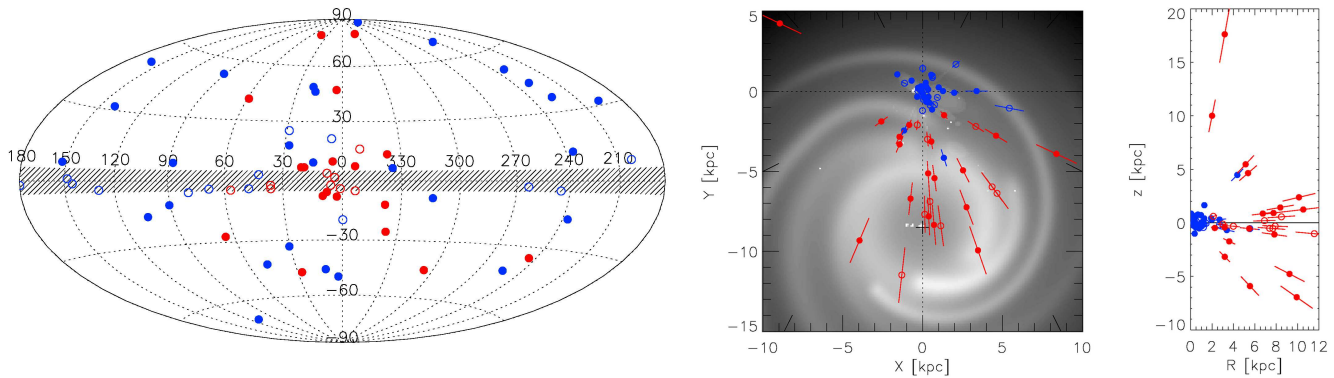}}  
\caption{Galactic distribution of the isolated pulsars from Table
  \ref{psr.table} (shown as blue circles) and pulsars in globular clusters
  from Table \ref{psr_gc.table} (red circles). Filled circles
  indicate the 45 lines of sight that we selected in
  \S~\ref{ss-los-selection}, the open circles indicate the
  remaining 23 lines of sight.  In the left panel we show the sight
  directions in a Hammer-Aitoff equal area projection, the hatched
  band indicates the region $|b|$ $<$ 5\degr. The middle panel shows the
  positions of the pulsars and globular clusters projected onto the
  Galactic mid-plane; the background image shows on a linear grey scale the column densities
  from the NE2001 model that were integrated perpendicular to the
  mid-plane. In the right panel we show the
  distribution of the pulsars in the $R_{\sun}, z$-plane, i.e. in the
  meridional planes through the Sun.  }
\label{los_distribution.fig}
\end{figure*}

\section{Introduction}
Knowledge of the distribution of the free electrons in the Galaxy,
$n_{\mathrm{e}}$, is important for several reasons. First, the
detailed structure of $n_{\mathrm{e}}$ is an essential ingredient for
a full understanding of the multi-phase character of the Galactic
interstellar medium (ISM). Furthermore, a model for the distribution of
the thermal electrons is a crucial tool for estimating the distances
of pulsars from their observed dispersion measure (DM)\footnote{The
  dispersion measure is the integral of the free electron density
  along the line of sight, and is expressed in units of
  cm$^{-3}$pc.}. Finally,
through its integral DM, an $n_{\mathrm{e}}$-model is required for an
analysis of the strength and structure of the Galactic magnetic field
from rotation measures (RM), because RM is the integral over the line
of sight of the product of the l.o.s.-component of the magnetic field
and $n_{\mathrm{e}}$ \citep[see, e.g.,][]{han2006, brown2007, men2008,
  noutsos2008, roy2008, vallee2008, nota2010}.


In our paper we analyse 8 models of the Galactic free electron density $n_{\mathrm{e}}$ that have appeared in the literature. These models range from single exponential thick disks, see, e.g., \citet{berkhuijsen2006},
\citet{berkhuijsen2008} and \citet{gaensler2008}, models that consist of a thin and a thick disk with different radial and vertical scale heights \citep{gomez2001}, and the complex, multi-component models by \citet{taylor1993}, or `TC93', and \citet{cordes2002}, also known as NE2001, that consist of an axisymmetric thin and thick disk, spiral arms, and a contribution by the Gum nebula (and, in the case of NE2001 additional components). We add four models to this list. The first model uses an exponential thick disk that is fitted to the latest available data on pulsars with accurate distances. We used a similar approach to fit new scale heights of the thick disk component in TC93 and NE2001, whilst maintaining the contributions by the other components in these models.  With our final model we consider
descriptions in which the prediction of DM is based on another
observable quantity, viz. emission measure (EM)\footnote{The emission
  measure is defined as the line-of-sight integral of the square of
  the free electron density, and is expressed in units of
  cm$^{-6}$pc.}. We then test the predictive quality of the available
models using a new formalism that we developed.

We arranged this paper as follows. We present and discuss the data in
Sect. \ref{s-data}, and in Sect. \ref{s-model} we give an overview of
the different models that we test. In Sect. \ref{s-details} we
describe the practicalities of modelling DM, in particular of the four
models we propose, and our criteria for selecting usable sightlines.
We then test the predictive quality of the available models in
Sect. \ref{s-comp}.
Finally, in Sect.~\ref{s-anomalousn} we discuss sightlines for which the DMs prove very difficult to predict. Throughout this
paper we have calculated statistics in a way that is robust against
outliers.

%
%

\section{The data}\label{s-data}
Our analysis is based on a sample of pulsars with relatively
accurate distances, either from a parallax measurement, or from an
association with a Galactic globular cluster. In our analysis we
selected lines of sight for which the fractional error in the parallax (isolated pulsars) or distance (pulsars in globular clusters) is less than 0.25; a similar approach was adopted by
\citet{berkhuijsen2008} and \citet{gaensler2008}. Small distance
errors make it easier to identify deviations between the modelled and
observed DMs, which are the result of structure in the ISM that is not
included in the model. 

Our sample consists of the (isolated) pulsars with known parallaxes
that are listed on Shami Chatterjee's
website\footnote{http://astrosun2.astro.cornell.edu/research/parallax/}
(accessed on 27/7/2012), for which we took the positions and measured
DMs from the ATNF pulsar catalogue \citep[version 1.44;][and this
website\footnote{http://www.atnf.csiro.au/research/pulsar/psrcat/}]{manchester2005}.
For all these pulsars we used the parallaxes that \citet{verbiest2010}
corrected for Lutz-Kelker bias. There are no DMs for 3 of the pulsars
listed in the ATNF pulsar catalogue (viz., J0633+1746, J0720-3125,
J1856-3754).
\citet{freire2010} published a parallax measurement for J1738+0333
that was not yet in the list on Chatterjee's website when we accessed
it, but we included it in our analysis. We corrected the parallax of J1738+0333 for Lutz-Kelker bias using the online tool at http://psrpop.phys.wvu.edu/LKbias/ . This selection results in a
list of 41 pulsars with known DMs and parallaxes, where we counted the
double pulsar J0737-3039 only once. The relevant properties of these
pulsars are listed in Table \ref{psr.table}.  In the case of
asymmetric parallax errors we made the errors symmetric by always
using the highest of the two error estimates. For the 68 sightlines which
we use in our analysis the
difference between the upper and lower error limits amounts to at most
4\%, and in the majority of cases to $\lesssim$ 1\%, of the
bias-corrected parallax. Also, the percentual error in the DMs of these
sightlines is at most 2\%, which means that the error in DM is smaller
than the error in the parallax by about an order of magnitude.
%
\citet{verbiest2010} list an additional 14 pulsars with fractional
parallax errors larger than 0.25 (six of these are larger than
one). We did not include these 14 pulsars since adding them would
considerably reduce the overall accuracy of the new sample compared to
the old sample, and would increase the number of sightlines by only a
modest amount.

While the distribution of the parallax errors can be well approximated by a Gaussian distribution, the distribution of the distance = 1/ parallax is not Gaussian. It can be shown that distance follows the probability density function

\begin{eqnarray}
\rho(\mathrm{dist} ; \pi_\mathrm{obs}, \mathrm{err}_\pi)\ =\ \frac{1}{\sqrt{2\mathrm{PI}}\ \mathrm{err}_\pi\ \mathrm{dist}^2}\ \mathrm{e}^{-0.5 (\frac{1}{\mathrm{dist}} - \pi_\mathrm{obs})^2/\mathrm{err}_\pi^2}
\label{distance_pdf}
\end{eqnarray} 

\noindent where $\pi_\mathrm{obs}$ and err$_\pi$ are the mean and error of the parallax distribution. To avoid confusion with $\pi$ = parallax, we used `PI' to indicate the mathematical constant $\pi$. In particular, the most probable distance is not 1/$\pi_\mathrm{obs}$,  but 1/(0.5$\pi_\mathrm{obs}$ + 0.5$\sqrt{\pi_\mathrm{obs}^2 + 8 \mathrm{err}_\pi^2}$). The most probable distance is about 11\% smaller than 1/$\pi_\mathrm{obs}$ for the smallest $\pi_\mathrm{obs}$/err$_\pi$ = 4 from our sample, and it approaches 1/$\pi_\mathrm{obs}$ asymptotically for large $\pi_\mathrm{obs}$/err$_\pi$. The distance errors in Table.~\ref{psr.table} define the most compact 68\% confidence interval around the most probable distance. By using Lagrange multipliers it can be shown that this interval is bounded by distances $a$ and $b$ for which the probability density function at $a$ equals the probability density function at $b$.

\begin{table*}
\centering
\caption{Properties of the 41 pulsars in our sample with fractional parallax errors $<$ 0.25. For each pulsar we list its Equatorial and Galactic coordinates, its measured DM, and the distance from the Sun (with associated error) and height above the Galactic mid-plane that were derived from parallax measurements. $\star$: these distance errors define the most compact 68\% confidence interval around the most probable distance. We excluded pulsars from our analysis for the following reasons: $^a$: pulsar lies within 5\degr\ of the Galactic plane; $^{b}$: line of sight is affected by strong H$\alpha$ emission in its vicinity; $^c$: line of sight has a BRT\_OBJ flag in \citet{finkbeiner2003}. In \S~\ref{ss-los-selection} we describe why we leave out sightlines that have an `a',`b' or `c' flag, and we use the remaining 27 sightlines for our analysis. }
\begin{tabular}{lrrrrrrrrr}
\hline
Name & \multicolumn{1}{c}{RA$_{\mathrm{2000.0}}$} & \multicolumn{1}{c}{DEC$_{\mathrm{2000.0}}$} &  \multicolumn{1}{c}{$l$}    &   \multicolumn{1}{c}{$b$}   & \multicolumn{1}{c}{DM}            & \multicolumn{3}{c}{dist$^\star$ +err -err} & \multicolumn{1}{c}{$|z|$}\\ 
     &                      &                      & \multicolumn{1}{c}{[\degr]} & \multicolumn{1}{c}{[\degr]} & \multicolumn{1}{c}{[cm$^{-3}$pc]}  & \multicolumn{3}{c}{[pc]} & \multicolumn{1}{c}{[pc]} \\ \hline   
   J0034-0721 &   00:34:08.87 &  -07:21:53.41 &   110.42 &   -69.82 &     11.38 &   1060 &  +100 &    -84 &    995 \\
   J0139+5814$^a$ &   01:39:19.74 &  +58:14:31.82 &   129.22 &    -4.04 &     73.78 &   2642 &  +320 &   -258 &    186 \\
   J0332+5434$^a$ &   03:32:59.37 &  +54:34:43.57 &   145.00 &    -1.22 &     26.83 &   1047 &  +140 &   -111 &     22 \\
   J0358+5413$^a$ &   03:58:53.72 &  +54:13:13.73 &   148.19 &     0.81 &     57.14 &   1081 &  +242 &   -168 &     15 \\
   J0437-4715 &   04:37:15.88 &  -47:15:09.03 &   253.39 &   -41.96 &      2.64 &    156 &    +1 &     -1 &    105 \\
   J0454+5543 &   04:54:07.75 &  +55:43:41.44 &   152.62 &     7.55 &     14.50 &   1182 &   +75 &    -66 &    155 \\
   J0538+2817$^a$ &   05:38:25.06 &  +28:17:09.16 &   179.72 &    -1.69 &     39.57 &   1371 &  +287 &   -203 &     40 \\
   J0630-2834 &   06:30:49.40 &  -28:34:42.78 &   236.95 &   -16.76 &     34.47 &    333 &   +53 &    -40 &     96 \\
   J0659+1414$^b$ &   06:59:48.13 &  +14:14:21.50 &   201.11 &     8.26 &     13.98 &    294 &   +37 &    -29 &     42 \\
J0737-3039A/B$^a$ &   07:37:51.25 &  -30:39:40.71 &   245.24 &    -4.50 &     48.92 &   1173 &  +269 &   -185 &     92 \\
   J0814+7429 &   08:14:59.50 &  +74:29:05.70 &   140.00 &    31.62 &      6.12 &    434 &   +10 &     -9 &    228 \\
   J0820-1350 &   08:20:26.38 &  -13:50:55.86 &   235.89 &    12.59 &     40.94 &   1937 &  +165 &   -141 &    422 \\
   J0826+2637 &   08:26:51.38 &  +26:37:23.79 &   196.96 &    31.74 &     19.45 &    362 &  +113 &    -70 &    191 \\
   J0835-4510$^a$ &   08:35:20.61 &  -45:10:34.88 &   263.55 &    -2.79 &     67.99 &    284 &   +17 &    -15 &     14 \\
   J0922+0638 &   09:22:14.02 &  +06:38:23.30 &   225.42 &    36.39 &     27.27 &   1151 &  +213 &   -156 &    683 \\
   J0953+0755 &   09:53:09.31 &  +07:55:35.75 &   228.91 &    43.70 &      2.96 &    262 &    +5 &     -5 &    181 \\
   J1012+5307 &   10:12:33.43 &  +53:07:02.59 &   160.35 &    50.86 &      9.02 &    826 &  +267 &   -164 &    641 \\
   J1022+1001 &   10:22:58.00 &  +10:01:52.76 &   231.79 &    51.10 &     10.25 &    556 &  +118 &    -83 &    432 \\
   J1136+1551 &   11:36:03.25 &  +15:51:04.48 &   241.90 &    69.20 &      4.86 &    355 &   +23 &    -20 &    331 \\
   J1239+2453 &   12:39:40.46 &  +24:53:49.29 &   252.45 &    86.54 &      9.24 &    845 &   +70 &    -60 &    843 \\
   J1456-6843 &   14:56:00.16 &  -68:43:39.25 &   313.87 &    -8.54 &      8.60 &    458 &   +76 &    -57 &     68 \\
   J1509+5531 &   15:09:25.63 &  +55:31:32.39 &    91.33 &    52.29 &     19.61 &   2067 &  +138 &   -122 &   1635 \\
   J1537+1155 &   15:37:09.96 &  +11:55:55.55 &    19.85 &    48.34 &     11.61 &   1026 &   +56 &    -50 &    766 \\
   J1543+0929 &   15:43:38.83 &  +09:29:16.34 &    17.81 &    45.78 &     35.24 &   6066 &  +865 &   -675 &   4347 \\
   J1559-4438 &   15:59:41.53 &  -44:38:45.90 &   334.54 &     6.37 &     56.10 &   2495 &  +696 &   -451 &    277 \\
   J1643-1224$^b$ &   16:43:38.16 &  -12:24:58.74 &     5.67 &    21.22 &     62.41 &    486 &  +128 &    -85 &    176 \\
   J1713+0747$^c$ &   17:13:49.53 &  +07:47:37.53 &    28.75 &    25.22 &     15.99 &   1069 &   +61 &    -54 &    456 \\
   J1738+0333 &   17:38:53.96 &  +03:33:10.84 &    27.72 &    17.74 &     33.78 &   1339 &   +97 &    -85 &    408 \\
   J1744-1134 &   17:44:29.40 &  -11:34:54.66 &    14.79 &     9.18 &      3.14 &    415 &   +18 &    -17 &     66 \\
   J1857+0943$^a$ &   18:57:36.39 &  +09:43:17.28 &    42.29 &     3.06 &     13.30 &   1019 &  +288 &   -186 &     54 \\
   J1909-3744$^c$ &   19:09:47.44 &  -37:44:14.38 &   359.73 &   -19.60 &     10.39 &   1264 &   +33 &    -31 &    424 \\
   J1932+1059$^a$ &   19:32:13.95 &  +10:59:32.42 &    47.38 &    -3.88 &      3.18 &    362 &    +9 &     -9 &     24 \\
   J2018+2839$^a$ &   20:18:03.83 &  +28:39:54.21 &    68.10 &    -3.98 &     14.17 &    981 &  +109 &    -89 &     68 \\
   J2022+5154 &   20:22:49.87 &  +51:54:50.23 &    87.86 &     8.38 &     22.65 &   1927 &  +313 &   -237 &    281 \\
   J2048-1616 &   20:48:35.64 &  -16:16:44.55 &    30.51 &   -33.08 &     11.46 &    951 &   +28 &    -26 &    519 \\
   J2055+3630$^b$ &   20:55:31.35 &  +36:30:21.47 &    79.13 &    -5.59 &     97.31 &   5277 & +1050 &   -755 &    514 \\
   J2124-3358 &   21:24:43.85 &  -33:58:44.67 &    10.93 &   -45.44 &      4.60 &    340 &   +96 &    -62 &    242 \\
   J2144-3933 &   21:44:12.06 &  -39:33:56.89 &     2.79 &   -49.47 &      3.35 &    169 &   +19 &    -16 &    128 \\
   J2145-0750 &   21:45:50.46 &  -07:50:18.44 &    47.78 &   -42.08 &      9.00 &    620 &  +154 &   -104 &    416 \\
   J2157+4017 &   21:57:01.85 &  +40:17:45.99 &    90.49 &   -11.34 &     70.86 &   3195 &  +826 &   -549 &    628 \\
   J2313+4253 &   23:13:08.62 &  +42:53:13.04 &   104.41 &   -16.42 &     17.28 &   1075 &   +88 &    -76 &    304 \\
\hline
\end{tabular}
\label{psr.table}
\end{table*}

\begin{table*}
\centering
\caption{Properties of the 27 globular clusters in our sample that are known to contain at least one pulsar. For each line of sight we list its Equatorial and Galactic coordinates, its measured DM, the number of pulsar the globular cluster contains, and its distance from the Sun and height above the Galactic mid-plane. We excluded lines of sight for the following reasons: $^a$: line of sight lies within 5\degr\ of the Galactic plane; $^{b}$: line of sight is affected by strong H$\alpha$ emission in its vicinity; $^c$: line of sight has a BRT\_OBJ flag in \citet{finkbeiner2003}. In \S.~\ref{ss-los-selection} we describe why we leave out sightlines that have an `a',`b' or `c' flag, and we use the remaining 18 sightlines for our analysis.
}
\begin{tabular}{lrrrrrrrr}
\hline
Cluster name & \multicolumn{1}{c}{RA$_{\mathrm{2000.0}}$} & \multicolumn{1}{c}{DEC$_{\mathrm{2000.0}}$} &  \multicolumn{1}{c}{$l$}    &   \multicolumn{1}{c}{$b$}   & \multicolumn{1}{c}{DM} & \multicolumn{1}{c}{N$_{\mathrm{psr}}$} & \multicolumn{1}{c}{dist} & \multicolumn{1}{c}{$|z|$}\\ 
     &                      &                      & \multicolumn{1}{c}{[\degr]} & \multicolumn{1}{c}{[\degr]} & \multicolumn{1}{c}{[cm$^{-3}$pc]}  & & \multicolumn{1}{c}{[pc]} & \multicolumn{1}{c}{[pc]} \\ \hline   
     NGC 104 &    00:24:05.2 &     -72:04:51 &   305.90 &   -44.89 &     24.37 &     23 &   4500 &   3200 \\
    NGC 1851 &    05:14:06.8 &     -40:02:48 &   244.51 &   -35.03 &     52.15 &      1 &  12100 &   6900 \\
    NGC 5024 &    13:12:55.3 &     +18:10:06 &   332.96 &    79.76 &     24.00 &      1 &  17900 &  17600 \\
    NGC 5272 &    13:42:11.6 &     +28:22:38 &    42.22 &    78.71 &     26.37 &      4 &  10200 &  10000 \\
    NGC 5904 &    15:18:33.2 &     +02:04:52 &     3.86 &    46.80 &     29.49 &      5 &   7500 &   5500 \\
    NGC 5986 &    15:46:03.0 &     -37:47:11 &   337.02 &    13.27 &     92.17 &      1 &  10400 &   2400 \\
    NGC 6121$^{b}$ &    16:23:35.2 &     -26:31:33 &   350.97 &    15.97 &     62.86 &      1 &   2200 &    600 \\
    NGC 6205 &    16:41:41.2 &     +36:27:36 &    59.01 &    40.91 &     30.17 &      5 &   7100 &   4700 \\
    NGC 6266 &    17:01:12.8 &     -30:06:49 &   353.57 &     7.32 &    114.08 &      6 &   6800 &    900 \\
    NGC 6342 &    17:21:10.1 &     -19:35:15 &     4.90 &     9.72 &     71.00 &      1 &   8500 &   1400 \\
    NGC 6397 &    17:40:42.1 &     -53:40:28 &   338.17 &   -11.96 &     71.80 &      1 &   2300 &    500 \\
    Terzan 5$^a$ &    17:48:04.8 &     -24:46:45 &     3.84 &     1.69 &    237.82 &     34 &   6900 &    200 \\
    NGC 6440$^a$ &    17:48:52.7 &     -20:21:37 &     7.73 &     3.80 &    222.80 &      6 &   8500 &    600 \\
    NGC 6441$^{b,c}$ &    17:50:13.1 &     -37:03:05 &   353.53 &    -5.01 &    232.24 &      4 &  11600 &   1000 \\
    NGC 6517 &    18:01:50.5 &     -08:57:32 &    19.23 &     6.76 &    182.45 &      4 &  10600 &   1200 \\
    NGC 6522$^a$ &    18:03:34.1 &     -30:02:02 &     1.02 &    -3.93 &    193.03 &      3 &   7700 &    500 \\
    NGC 6539 &    18:04:49.7 &     -07:35:09 &    20.80 &     6.78 &    186.38 &      1 &   7800 &    900 \\
    NGC 6544$^a$ &    18:07:20.6 &     -24:59:50 &     5.84 &    -2.21 &    134.00 &      2 &   3000 &    100 \\
    NGC 6624 &    18:23:40.5 &     -30:21:40 &     2.79 &    -7.91 &     87.63 &      6 &   7900 &   1100 \\
    NGC 6626 &    18:24:32.8 &     -24:52:11 &     7.80 &    -5.58 &    120.10 &     12 &   5500 &    500 \\
    NGC 6656 &    18:36:23.9 &     -23:54:17 &     9.89 &    -7.55 &     91.40 &      2 &   3200 &    400 \\
    NGC 6749$^a$ &    19:05:15.3 &     +01:54:03 &    36.20 &    -2.21 &    192.85 &      2 &   7900 &    300 \\
    NGC 6752 &    19:10:52.1 &     -59:59:04 &   336.49 &   -25.63 &     33.36 &      5 &   4000 &   1700 \\
    NGC 6760$^a$ &    19:11:12.0 &     +01:01:50 &    36.11 &    -3.92 &    199.68 &      2 &   7400 &    500 \\
    NGC 6838$^a$ &    19:53:46.5 &     +18:46:45 &    56.75 &    -4.56 &    117.00 &      1 &   4000 &    300 \\
    NGC 7078 &    21:29:58.3 &     +12:10:01 &    65.01 &   -27.31 &     66.88 &      8 &  10400 &   4800 \\
    NGC 7099 &    21:40:22.1 &     -23:10:48 &    27.18 &   -46.84 &     25.08 &      2 &   8100 &   5900 \\ \hline
\end{tabular}
\label{psr_gc.table}
\end{table*}

Paulo Freire lists pulsars in globular clusters on his
website\footnote{http://www.mpifr-bonn.mpg.de/staff/pfreire/GCpsr.txt}
(accessed on 27/7/2012). We obtained the properties of their parent
globular clusters from \citet[][and this
website\footnote{http://physwww.mcmaster.ca/$\sim$harris/mwgc.dat},
version Dec. 2010]{harris1996}. We list the globular clusters with
their measured DM in table \ref{psr_gc.table}. When there is more
than one pulsar in a globular cluster we calculated the average DM
and its 1-$\sigma$ spread. Clusters that are rich in pulsars, like NGC
104 and Terzan 5, show a statistically significant spread in pulsar
DMs, which means that there is warm ionized gas inside globular
clusters. The spread in DMs is however small, only a few percent of
the average DM for each cluster, so that we can average the pulsar
DMs in each cluster, and use this DM as the characteristic value
for that line of sight. This yields DMs for 27 globular clusters with
known distances. Following \citet{gaensler2008} (who base their choice
on a private communication of C. Heinke), we adopt a fractional
uncertainty in the distance of each globular cluster of 15\%.

\citet{gaensler2008} include also pulsar DMs in the Magellanic Clouds
in their analysis as upper limits on the Galactic DMs for these
sightlines. These upper limits are much higher than the DMs we find for
objects far from the Galactic plane, which means that they must contain a
significant contribution from the Clouds themselves. Therefore we did 
not include these data in our analysis.

In Fig. \ref{los_distribution.fig} we show the distribution of the 68
lines of sight from Tables \ref{psr.table} and \ref{psr_gc.table} in 3
different projections. This figure shows that the sightlines towards
pulsars with parallax distances sample structure in the Galactic ISM
within a few kpc from the Sun, and the sightlines towards pulsars in
globular clusters sample longer sightlines that mostly lie further
than 3\degr -- 4\degr\ from the Galactic plane. In
\S~\ref{ss-los-selection} we describe our criteria for selecting 45
sightlines from this sample that we then subsequently use to determine the parameters for the $n_\mathrm{e}$ models.
In Sect.~\ref{s-comp} we analyse how well the different models predict DMs for the subset of 45 `good' lines of sight, and for the full sample of 68 sightlines.

\begin{table*}
\centering
\caption{Overview of the models that we considered, and the accuracy of the DMs that they predict. For each model we list the factor N by which the modelled DM deviates from the measured DM (see Sect.~\ref{s-comp}) for the 50th, 75th and 90th percentile of the ensemble of 45 selected lines of sight, and all 68 lines of sight, as we describe in Sect.~\ref{s-comp}. We used a linear interpolation between data points to calculate the N that correspond to exactly the 50th, 75th and 90th percentiles. $^\star$: N was not calculated for model M4 when all 68 sightlines are used, because the EM for some of these sightlines cannot be calculated reliably (see \S~\ref{ss-calculating-em}). 
References: (1) \citet{berkhuijsen2008}; (2) \citet{gaensler2008}; (3)  \citet{gomez2001}; (4) \citet{taylor1993}; (5) \citet{cordes2002}; (6) \citet{sun2010}; (7) \citet{law2011}; (8) \citet{jansson2012}. 
}
\begin{tabular}{llccccccc}
\hline
Model & Characteristics & Reference & \multicolumn{6}{c}{Predictive quality N for} \\ 
      &                 &           &  50\% & 75\% & 90\% &  50\% & 75\% & 90\% \\ 
      &                 &           & \multicolumn{3}{c}{of 45 lines of sight} & \multicolumn{3}{c}{of all lines of sight} \\ 
\multicolumn{6}{l}{\emph{Exponential models with a single scale height}} \\
BM08$^{\mathrm{a}}$        & $h$ = 0.93 kpc, DM$_{\infty,\perp}$ = 21.7 cm$^{-3}$pc & 1 & 1.33 & 1.48 & 1.74 & 1.41 & 1.65 & 2.28 \\
GMCM08$^{\mathrm{b}}$ & $h$ = 1.83 kpc, DM$_{\infty,\perp} $ = 25.6 cm$^{-3}$pc & 2 & 1.20 & 1.70 & 2.12 & 1.49 & 2.09 & 3.13 \\
M1$^{\mathrm{c}}$             & $h$ = 1.59 kpc, DM$_{\infty,\perp}$ = 24.4 cm$^{-3}$pc  & \emph{this paper} & 1.21 & 1.60 & 2.03 & 1.39 & 1.94 & 2.86 \\ \\
\multicolumn{4}{l}{\emph{Plane-parallel 2-component models with a radial scale length}} \\
GBCa                & Eqn. \ref{rho_gomez.eqn} with  $f(x)\ =\ \mathrm{e}^{-x}       $  & 3 & 1.32 & 1.46 & 1.72 & 1.36 & 1.63 & 2.17 \\
GBCb                & Eqn. \ref{rho_gomez.eqn} with  $f(x)\ =\ \mathrm{sech}^2(x)$  & 3 & 1.35 & 1.48 & 1.68 & 1.37 & 1.60 & 2.15 \\ \\
\multicolumn{4}{l}{\emph{Multi-component models}} \\
TC93                 & Model by Taylor \& Cordes                                                            & 4 & 1.33 & 1.50 & 1.81 & 1.35 & 1.56 & 2.33 \\
M2$^{\mathrm{d}}$  & TC93 with a thick disk of 1.59 kpc         & \emph{this paper} & 1.20 & 1.37 & 1.82 & 1.22 & 1.53 & 2.03 \\
NE2001a             & Model by Cordes \& Lazio (Web interface)                             & 5 & 1.26 & 1.79 & 2.29 & 1.29 & 1.86 & 2.86 \\ 
NE2001b             & Model by Cordes \& Lazio (Fortran routine)                           & 5 & 1.25 & 1.76 & 2.25 & 1.27 & 1.77 & 2.85 \\
NE2001c             & Model by Cordes \& Lazio                                                   & 6,7,8 & 1.25 & 1.52 & 1.85 & 1.32 & 1.62 & 2.04 \\ 
                              & (Fortran routine + thick disk from GMCM08)   & \\
M3$^{\mathrm{d}}$     & NE2001b with a thick disk of  1.31 kpc, & \emph{this paper} & 1.29 & 1.51 & 1.83 & 1.35 & 1.59 & 2.02 \\
                    & DM$_{\infty,\perp}$= 20.5 cm$^{-3}$pc  \\  \\
\multicolumn{4}{l}{\emph{Structured model using DM(EM)}} \\
M4$^{\mathrm{d}}$     & DM$_\infty=22.67\ \mathrm{EM}_\infty^{0.64}$ + M2 for extrapolation & \emph{this paper} & 1.98 & 2.53 & 3.24 & $^\star$ & $^\star$ & $^\star$ \\
                    & to finite distance \\ 
\hline
\end{tabular}
\begin{list}{}{}
\item[$^{\mathrm{a}}$]: constrained by 38 lines of sight, err$_{\mathrm{dist}}$/dist $<$ 0.5 , effective $\tau_{\mathrm{H}\alpha}$ $<$ 1
\item[$^{\mathrm{b}}$]: constrained by 16 lines of sight with $|b|$ $>$ 40\degr, err$_{\mathrm{dist}}$/dist $<$ 0.33
\item[$^{\mathrm{c}}$]: constrained by 22 lines of sight from Tables \ref{psr.table} and \ref{psr_gc.table} with $|b|$ $>$ 32.5\degr, err$_{\mathrm{dist}}$/dist $<$ 0.25, effective $\tau_{\mathrm{H}\alpha}$ $<$ 1.
\item[$^{\mathrm{d}}$]: constrained by 44 lines of sight from Tables \ref{psr.table} and \ref{psr_gc.table} with $|b|$ $>$ 5\degr, err$_{\mathrm{dist}}$/dist $<$ 0.25, effective $\tau_{\mathrm{H}\alpha}$ $<$ 1.
\end{list}
\label{DMmodels.table}
\end{table*}

\section{Overview of electron density models}\label{s-model}
In this section we give an overview of the models that were published in the literature, supplemented with a description of the models M1-M4 that we introduced. Table~\ref{DMmodels.table} summarizes the inner workings of the
$n_{\mathrm{e}}$-models for which we will analyse the quality of the
DM predictions. In our overview we start with the simplest imaginable models, consisting of a single exponential thick disk, and build towards the multi-components models by \citet{taylor1993} and \citet{cordes2002}.

\subsection{Plane-parallel exponential models with a single scale height, model M1}\label{ss-expfixed}
\citet{berkhuijsen2008} and \citet{gaensler2008} have investigated
models with a plane-parallel $n_{\mathrm{e}}$-distribution with a single scale
height. I.e., 
\begin{eqnarray}
n_{\mathrm{e}}(x,y,z) \equiv n_{\mathrm{e}}(z) = n_0 \mathrm{e}^{- |z|/h}
\end{eqnarray} 
\noindent
so that the dispersion measure 
\begin{eqnarray}
DM(L,b) \equiv \int_{0}^{L} n_{\mathrm{e}}\ dl\ =\ \frac{n_0 h}{\sin(|b|)} [1 - \mathrm{e}^{- L\sin(|b|) / h}]
\label{finiteDM.eqn}
\end{eqnarray}
\noindent 
where $L$ is the length of the line of sight, and $b$ is its Galactic
latitude. The fraction on the right hand side is the DM of an
infinitely long line of sight, DM$_\infty$, and the term between
brackets corrects DM$_\infty$ for a finite path length.
For a uniform free electron density in the Galactic mid-plane the DM
that is accumulated perpendicular to the disk can be easily calculated
as DM$_\perp$ = DM$\sin(|b|)$. In this paper we will often refer to
this quantity, and to the DM$_\perp$ of an infinitely long line of
sight, DM$_{\infty,\perp}=n_0 h$. \cite{berkhuijsen2006}, \cite{berkhuijsen2008}, and
\cite{gaensler2008} find somewhat different DM$_{\infty,\perp}$ of 25, 22, and 26
cm$^{-3}$pc resp., which can be related to the different
criteria that these authors used to select lines of sight for their
analysis (we list these in Table \ref{DMmodels.table}).

Our model M1 consists of only a single exponential thick disk, similar to the models by Berkhuijsen et al. and Gaensler et al., for which we determine the best scale height in \S~\ref{ss-model-m1-details} based on the data from Tables~\ref{psr.table} and \ref{psr_gc.table}.

%




\subsection{Plane-parallel 2-component models with a radial scale length}\label{ss-gomez}

\citet{gomez2001} considered models that consist of a thin and thick disk that each show also a dependence on Galactocentric
distance. They determined the parameters of these models using a data set of 109 pulsars.  (Throughout our paper, when we use the term `Galactocentric radius' or `Galactocentric distance' we are referring to a distance projected onto the Galactic mid-plane) For only 8 of their
pulsars a parallax distance was available, and for 27 of their pulsars
the distance is based on an association with either a globular cluster
or a supernova remnant. Most of the remaining distances are
kinematic. The errors in the latter were discussed by
\citet{gomez2006} and were found to be significant, especially at the
positions of the spiral arms.
The general form of the electron density from which \citet{gomez2001} calculated DMs is
\begin{eqnarray}
n_{\mathrm{e}}(r_{\mathrm{GC}},z) = n_1 \frac{f(r_{\mathrm{GC}}/h_{r,1})}{f(r_{\sun}/h_{r,1})}f(z/h_{z,1})\ +\ n_2 \frac{f(r_{\mathrm{GC}}/h_{r,2})}{f(r_{\sun}/h_{r,2})}f(z/h_{z,2})
\label{rho_gomez.eqn}
\end{eqnarray}
where $z$ is the height above the Galactic mid-plane, $z~=~L\sin(|b|)$, and $f(x)$ = $\mathrm{e}^{-x}$ (which we refer to in
Table \ref{DMmodels.table} as `GBCa') or $f(x) = \mathrm{sech}^2$(x)
(`GBCb').  In Table~\ref{Gomez_parameters.table} we list the fit
parameters obtained by \citet{gomez2001}.  Their model is somewhat
reminiscent of that by \citet{lyne1985}, which consisted of a thick
and thin disk component (the latter having a small vertical scale
height). The thick disk in the model by \citet{lyne1985} provides a
constant free electron density, contrary to the other models that we
tested.

\begin{table}
\centering
\caption{Parameters of the models GBCa (which uses f(x)=exp(-x)) and GBCb (f(x)=$\mathrm{sech}^2$(x)) that give the best fit to the data from \citet{gomez2001}.}
\begin{tabular}{llrrr}
\hline
Model & f(x) & n(r=r$_{\sun}$,z=0) & $h_z$ & $h_r$ \\
      &      &   [cm$^{-3}$]       & [kpc] & [kpc] \\ \hline
GBCa  & exp(-x)     & 2.03 $\times$ 10$^{-2}$ & 1.07 & 30.4 \\
      &      & 0.71 $\times$ 10$^{-2}$ & 0.05 &  1.5 \\
GBCb  & sech$^2$(x) & 1.77 $\times$ 10$^{-2}$ & 1.10 & 15.4 \\
      &      & 1.07 $\times$ 10$^{-2}$ & 0.04 &  3.6 \\
\hline
\end{tabular}
\label{Gomez_parameters.table}
\end{table}


\subsection{Multi-component models TC93 and NE2001a,b,c, models M2 and M3}\label{ss-CL02}

This family of models is built around the same basic structure for
the Milky Way, which was first described by \citet{taylor1993}. Their
model consists of a thick disk with a large radial scale length and a
large extent perpendicular to the mid-plane, an additional thin disk
component close to the Galactic plane, 4 spiral arms whose locations
are based on the model by \citet{georgelin1976} that were subsequently
modified slightly, and an enhancement of the free electron density in
the Gum nebula. The models by \citet{lyne1985} and \citet{cordes1991}
contained only thin disk and thick disk components, but
\citet{ghosh1992} and \citet{taylor1993} argued the need for spiral
arms. In the remainder we will refer to the model by
\citet{taylor1993} as `TC93'.
%

The model that was defined by \citet{cordes2002} consists of 7
components, each with its analytical expression for the free electron
density $n_{\mathrm{e}}$ as a function of position in the
Galaxy. There are 3 smooth components: the spiral arms and the thin
and thick disk. Together, these are defined by 28 parameters. Then
there is a Galactic-centre component (defined by 3 parameters), and a
description of the Local ISM (which requires 36 parameters). Finally,
there is a significant number of clumps and voids with which the
authors describe the smaller-scale structure in the ISM. 
%

We compare the DM-predictions of two versions of the NE2001
model with the data in Tables~\ref{psr.table} and
\ref{psr_gc.table}. The first version is defined by the web
interface\footnote{http://rsd-www.nrl.navy.mil/7213/lazio/ne$\_$model/}
(NE2001a); the second one is defined by the Fortran code that can be
downloaded from the same web address (NE2001b). In Sect.~\ref{s-comp} we will show that the web interface and Fortran code predict different DMs for the same sightlines. Three papers have recently been published \citep{sun2010,law2011,jansson2012} in
which the authors replaced the scale height of the thick disk in the
NE2001 model (which was $h$ = 970 pc\footnote{The Fortran code uses
  970 pc, but Table 3 in \citet{cordes2002} lists 950 pc. We will
  adopt 970 pc as the correct value.}) with the much larger value of
$h$ = 1830 pc that was found by \citet{gaensler2008}. The mid-plane
density was also replaced by the value used by \citet{gaensler2008},
which changes the DM$_{\infty,\perp}$ of the thick disk from 33
cm$^{-3}$pc down to 26 cm$^{-3}$pc. We will refer to this as model
NE2001c. 

With our models M2 and M3 we investigate whether changing the scale height of the thick disk in TC93 resp. NE2001b improves their accuracy in predicting DMs. (\S~\ref{ss-model-m23-details})

\subsection{M4: A structured model based on the DM(EM)-relation}\label{ss-structured}

The strength of the models we discussed in the previous
subsections is that they require only a small number of model components. However, on small scales the ISM is not as smooth as these models predict. With our model M4 we test whether
one can predict DM more accurately by using the emission measure EM as
a proxy to predict DM$_\infty$, and then use one of the other models to correct for the finite distance to the pulsar. This approach could in principle outperform other models, since
the EM can reflect the structure in DM on small physical scales. In \S~\ref{ss-model-m4-details}, we will test whether model
M4 indeed gives a smaller scatter in DM$_{\infty,\perp}$ than other models.

One caveat in this approach is that the same electrons are assumed to
produce both DM and EM. \citet{snowden1997} showed by analysing
ROSAT X-ray data that the mid-plane density of free electrons in the
Hot inter-cloud medium (HIM) is 3.5 $\times$ 10$^{-4}$ cm$^{-3}$, which
is an order of magnitude smaller than the mid-plane density of the free
electrons in the Warm ionized medium (WIM). We therefore neglected
their contribution to DM, and, as a consequence, both the observed
DM and EM are produced only by electrons in the WIM.

\section{Details of the investigated models}\label{s-details}

In this section we describe the practicalities of calculating the
model DMs.  We will first describe how we calculate EM, since we will use EM to remove sightlines in the direction of complex H$\alpha$ features. In \S~\ref{ss-los-selection} we
then describe how we selected 45 `clean' lines of sight from Tables
\ref{psr.table} and \ref{psr_gc.table} that we will use to fit model parameters.  Finally, in
\S~\ref{ss-model-m1-details} to \ref{ss-model-m4-details} we will
discuss the inner workings of models M1--M4 in more detail. In \S~\ref{ss-model-m23-details} we will also explain why the thick disk from NE2001 cannot be exchanged for the thick disk from GMCM08.

For the models from the literature (TC93, NE2001a,b,c, BM08, GMCM08,
GBCa and GBCb), we took the parameters such as vertical scale height
$h$, mid-plane density $n_0$, and radial scale length (for GBCa and
GBCb) from the respective papers, to predict DMs for each of the 45
lines of sight from Tables \ref{psr.table} and \ref{psr_gc.table}. 

\subsection{Calculating EM}\label{ss-calculating-em}

The EMs are calculated in two steps. In the first step, the observed
H$\alpha$ intensities are translated into `observed' EM, or
EM$_{\mathrm{obs}}$. In the second step these are converted to EM for
the entire line of sight, by correcting for interstellar extinction.

The first step in this calculation starts with the all-sky map
produced by \citet{finkbeiner2003} from the WHAM, SHASSA and VTSS
surveys (see \citet{haffner2003}, \citet{gaustad2001} and
\citet{dennison1998} resp.). The required conversion factor is quite
well established; see, e.g., \citet{haffner1998}, and the detailed
discussion in \citet{vallsgabaud1998}, in particular his `Case B'
scenario of optically thick Lyman continuum radiation. It appears that
different authors agree on the conversion factor
EM$_{\mathrm{obs}}/I_{\mathrm{H}\alpha}$ to within 2\% for temperatures around
8000 to 9000 K, which are usually assumed for this ISM phase
\citep{reynolds1985,madsen2006}. We have therefore used the relation
given by \citet{haffner1998}, viz.: EM$_{\mathrm{obs}}\ =\ 2.75\
{(T/10^4)}^{0.9} I_{\mathrm{H}\alpha}$, with EM in the usual units,
$I_{\mathrm{H}\alpha}$ in Rayleigh, and $T$ = 8000 K.

The second step, i.e. the correction for extinction, is less direct.
The correction is $\mathrm{e}^\tau$, where $\tau$ is the effective
optical depth of the H$\alpha$-emission. The total optical depth
follows from the $E(B-V)$ that \citet{schlegel1998} derived, assuming
$R = A_V/E(B-V) =3.1$, and taking $A_{H\alpha}/A_V = 0.81$
\citep[see][and references therein]{dickinson2003}. The effective
optical depth depends on the relative distributions of the diffuse gas
and the dust along the line of sight; \citet{dickinson2003} derived
that the effective optical depth of the H$\alpha$ line is 0.33 $\pm$
0.10 times the total optical depth.  The correction factor for
interstellar extinction then becomes $\mathrm{e}^{0.76\ \times\
  \mathrm{E(B-V)}}$. We did not use the values of $A_V$ given by
\citet{rowles2009}, because their results are quite comparable to
those of Schlegel et al. for $|b|$ $>$ 5\degr, while \citet{peek2010}
confirm the Schlegel et al. $E(B-V)$-values to within 0.05 magnitudes,
although the maps that Peek et al. made have a resolution of only
4.5\degr.

We limit the analysis to pulsars with $|b|$ $>$ 5\degr, because
\citet{schlegel1998} warn that the $E(B-V)$ they derive are only
reliable for these latitudes. Also, \citet{dickinson2003} and
\citet{berkhuijsen2008} point out that one can only reliably correct
for interstellar extinction when the effective optical depth is
smaller than 1, which excludes most lines of sight with $|b|$ $<$
5\degr.

There is an additional complication when estimating EM for pulsars
in globular clusters. Globular clusters are bright in H$\alpha$, but most of 
this emission is produced by the stars in the cluster, not by the
gas. We can therefore not use the H$\alpha$ intensity of the cluster
itself to calculate the Galactic EM along the line of
sight. Instead, we calculate the average EM for a 3$\times$3
rectangular grid of lines of sight, excluding the central line of
sight. The cells in this grid measure 18\arcmin$\times$18\arcmin,
which puts the outer lines of sight outside the edge of the largest
globular clusters in our sample.

\subsection{Selection of the lines of sight}\label{ss-los-selection}

We selected a subset of 27 lines of sight from Table \ref{psr.table}
and 18 lines of sight from Table \ref{psr_gc.table} to determine the fit parameters for models M1--M4 from Table~\ref{DMmodels.table}.
First we left out all lines of sight
that lie within 5\degr\ of the Galactic plane. This selection avoids regions
close to the Galactic plane where structure in the ISM generally
becomes very complicated, and difficult to describe without detailed
modelling. Also, as we described in
\S~\ref{ss-calculating-em}, this avoids lines of sight with unreliable
EM where our model M4 cannot be used.

For the remaining lines of sight we added 2 more selection
criteria. The high-resolution H$\alpha$ maps from VTSS and SHASSA
showed strong, structured H$\alpha$ emission in the direction of the
pulsars J0659+1414, J1643-1224, and J2055+3630, and the 2 globular
clusters NGC 6121 and NGC 6441. The lines of sight towards J1713+0747,
J1909-3744, and NGC 6441 furthermore have `BRT\_OBJ' flags in the
H$\alpha$ mask that was created by \citet{finkbeiner2003}, which
indicates regions where the automated removal of bright point sources
cannot be trusted. Since both effects produce unreliable EM we
excluded these lines from our sample (NGC 6441 is removed for both
these reasons). Finally, we checked that the remaining lines of sight
all have an effective optical depth $<$ 1, which is required to
reliably correct for interstellar extinction, as described in \S~\ref{ss-calculating-em}.

The remaining sample of lines of sight that we will use to determine the parameters of the models M1--M4 contains 27 lines of sight towards
isolated pulsars, and 18 lines of sight towards pulsars in globular
clusters.  As a consequence of our choice to use relatively
unsophisticated selection criteria, a few of the selected lines of
sight will have DMs that are difficult to model, as will become clear
in the remainder of this paper. We will return to this in
Sects.~\ref{s-comp} and \ref{s-anomalousn}.

\begin{figure}
\resizebox{\hsize}{!}{\includegraphics{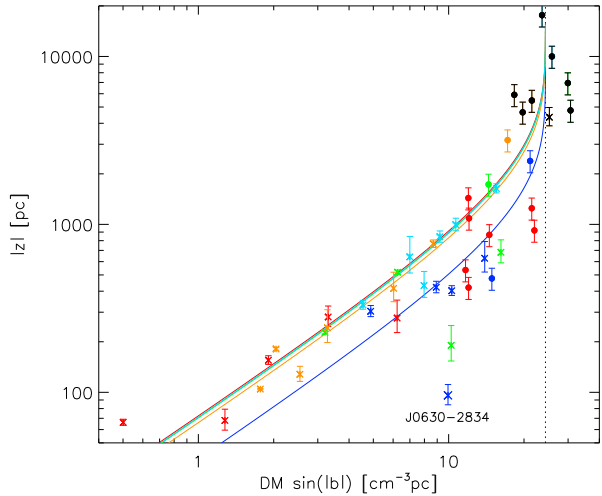}}
\caption{Distribution of height above the Galactic mid-plane $|z|$
  vs. DM$_\perp=DM\sin(|b|)$ for selected pulsars from
  Table \ref{psr.table} and globular clusters from Table
  \ref{psr_gc.table} (crosses and filled circles resp.). The dotted
  line indicates the asymptotic value for DM$\sin(|b|)\ =\ 24.4\
  \mathrm{cm}^{-3}$pc that we calculated from all lines of sight with
  $|z|\ >\ $ 4 kpc, which are indicated with black symbols. Colours
  identify lines of sight that belong to the same latitude bin, and
  the coloured lines represent the best fits to the points with the
  same colour. The 5 colours shown in this Figure correspond to the
  data points that are drawn with solid error bars in
  Fig. \ref{h_n0.fig}; red belongs to the latitude bin 5\degr\ $< |b|<
  $ 10\degr, dark blue to the next latitude bin, etc. We indicate the
  pulsar J0630-2834 that we did not use to determine $h$, but that we
  did use in our analysis from
  Sect. \ref{s-comp}.   }
\label{dmperp_z.fig}
\end{figure}

\begin{figure}
\resizebox{\hsize}{!}{\includegraphics{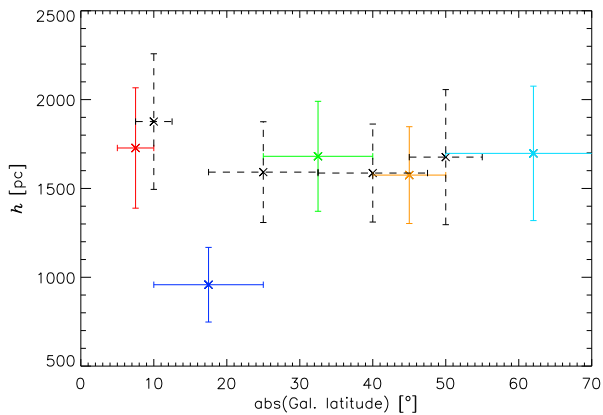}}
\caption{The distribution of fitted scale heights $h$ as a function of Galactic latitude. The vertical error bars reflect the 1-$\sigma$ spread in the derived $h$, and the horizontal error bars indicate the latitude range $\Delta b$ of the lines of sight that are used in the fit. The rightmost data point covers all latitudes $|b|$ $>$ 50\degr, and is centred on the average $b$ of this latitude range, weighted by the surface area of each annulus d$b$. With this choice of bin widths there are at least 5 lines of sight per latitude bin. Points with dashed lines use the same $\Delta b$ as the data points immediately to the left of it.  
}
\label{h_n0.fig}
\end{figure}

\subsection{Parameters for model M1}\label{ss-model-m1-details}
As described in Sect.~\ref{s-data}, distances derived from parallaxes follow a complicated, non-Gaussian error distribution, which needs to be taken into account properly when fitting a scale height to these data points.  We find the best-fitting scale height by maximising the likelihood $\mathcal{L}$, where each pulsar with a parallax measurement contributes a probability density for the distances given by Eqn.~\ref{distance_pdf}, and each globular cluster with a DM measurement contributes a Gaussian distribution of the distances. The log(likelihood) is given by

\begin{eqnarray}
\mathrm{log}(\mathcal{L})\ \propto\ \sum_\mathrm{pulsars} -\frac{1}{2}\frac{\left( \pi - \frac{1}{L\_\mathrm{pred(h)}}\right)^2}{\mathrm{err}\_\pi^2} -2\ \mathrm{log}(L\_\mathrm{pred(h)}) \nonumber \\
+\ \sum_\mathrm{clusters} -\frac{1}{2}\frac{\left( L - L\_\mathrm{pred(h)}\right)^2}{\mathrm{err}\_L^2}
\label{ml_criterion}
\end{eqnarray}  

\noindent where $\pi$ and $L$ are used as shorthand for the parallax and distance of the pulsar/globular cluster resp., $L\_\mathrm{pred(h)}$ indicates the distance of the pulsar/globular cluster that is predicted from its observed DM and input scale height $h$, and `err' stands for the measurement error. We numerically evaluated  Eqn.~\ref{ml_criterion} for a grid of scale heights, and we fitted a parabola to the maximum of the distribution of  $\mathrm{log}(\mathcal{L})$ as a function of scale height to find the overall best-fitting scale height. 
We used this technique to determine the optimum scale height for all three models M1--M3 that we introduced.

GCMC08 previously showed that the value of the scale height depends on which sightlines are used to fit the exponential thick disk to. While they changed the minimum $|b|$ to find the best-fitting scale height for the thick disk, we divide the sightlines over bins in Galactic latitude and analyse how the fitted scale heights differ between the latitude bins, see Figs.~\ref{dmperp_z.fig} and \ref{h_n0.fig}. 

\begin{figure*}
\resizebox{\hsize}{!}{\includegraphics{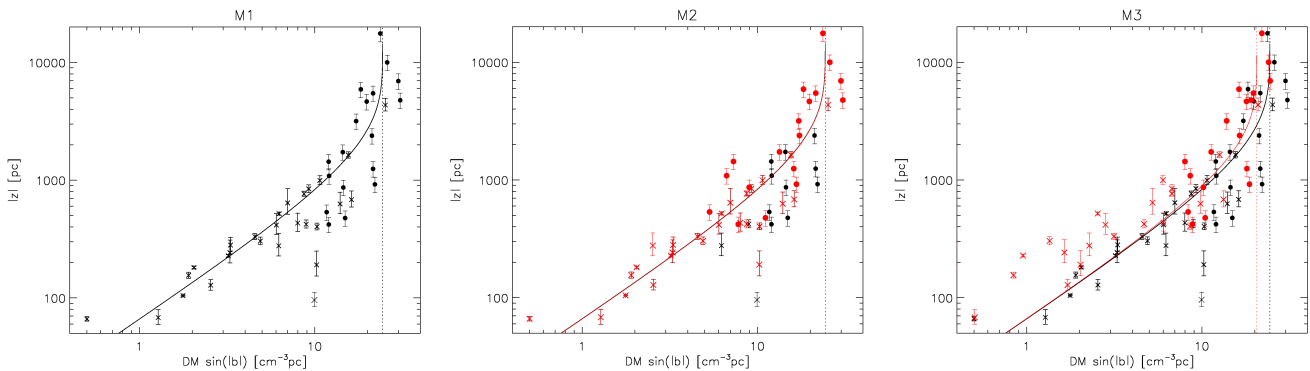}}
\caption{Comparison between the thick disk components in models M1--M3 and the data these components were fitted to. Crosses indicate isolated pulsars, filled circles indicate pulsars in globular clusters. 
Black data points show the original DM$\sin(|b|)$ and $|z|$ data points from Tables~\ref{psr.table} and \ref{psr_gc.table}, and the black model curve shows the exponential disk model M1 that was fit to these data. Red data points show the DM$\sin(|b|)$ residuals (explained in \S~\ref{ss-model-m23-details}) and model curves for the exponential disk components in models M2 and M3 that were fitted to the red data points. The model curves asymptotically approach the vertical dotted line, which is determined from the DM$\sin(|b|)$ (model M1) or residual DM$\sin(|b|)$ (M2,M3) of target sources at $|z|$ $>$ 4 kpc.
}
\label{z_dmsinb_m123.fig}
\end{figure*}

First, when fitting a scale height, we solve for the height above the Galactic midplane $z$ as a function of DM$_\perp$ instead of the other way round, since the errors in $z$ are larger than the errors in DM$_\perp$ by about an order of magnitude (or more). As \citet{savage2009} pointed out this can lead to a biased estimate of the fitted scale height, and we will return to this in the final paragraph of this section. The curves that we fitted to the data points are required to converge asymptotically to the DM$_\perp$ at infinity, DM$_{\infty,\perp}$, in all latitude bins. We approximated DM$_{\infty,\perp}$ by the DM$_\perp$ of the 8 lines of sight with $|z|$ $>$ 4~kpc, on the assumption that very little DM$_{\perp}$ is built up beyond those $|z|$. These 8 data points consist of 7 globular clusters and 1 pulsar that all lie at least 27\degr\ from the Galactic plane, and are indicated in black in Fig.~\ref{dmperp_z.fig}. This way we find DM$_{\infty,\perp}$ = 24.4 $\pm$ 4.2 cm$^{-3}$pc. We then determined the scale height of the exponential thick disk for each latitude bin by maximising $\mathrm{log}(\mathcal{L})$ in the way we described in the first paragraph of this section. 

We colour-coded sightlines that belong to the same latitude bin, and the colours in Fig.~\ref{dmperp_z.fig} match the colours in Fig.~\ref{h_n0.fig}. 
We used a Monte Carlo simulation with 10$^4$ draws to simulate the effect that the scatter in DM$_{\infty,\perp}$ has on the fitted scale heights, and we will use the same approach to determine the uncertainty in the fitted scale heights in models M1--M3. The horizontal error bars in Fig.~\ref{h_n0.fig} indicate the range that is covered by each bin in Galactic latitude, the widths of these bins were chosen in such a way that each bin contains at least 5 sightlines that we can fit  a scale height with. Data points with dashed error bars in Fig.~\ref{h_n0.fig} show the fit results when the bins are shifted further off the Galactic plane by half a binwidth; adding these bins better samples the fitted scale height $h$ as a function of Galactic latitude. 

As Fig.~\ref{h_n0.fig} shows, the scale heights that are fitted to data points in the higher latitude bins are consistent with each other, only the scale height fitted to the blue data points in Fig.~\ref{dmperp_z.fig} (sightlines with 10\degr $<$ $|b|$ $<$ 25\degr) jumps out. When fitting the scale height of the exponential thick disk in our model M1 we stayed clear of this region by conservatively selecting only sightlines with $|b|$ $>$ 32.5\degr; this selection also avoids sightlines between 17.5\degr $<$ $|b|$ $<$ 32.5\degr\  where the transition to the smaller scale height that was fitted to the blue data points occurs. We found a scale height $h$ = 1586 $\pm$ 295 pc for the thick disk in M1.


\citet{savage2009} point out in the Appendix of their paper that an
exponential model cannot reproduce observed DM$_\perp$ that are larger
than the DM$_{\infty,\perp}$ of the model, and that this introduces a
bias in the derived scale height of the model. They conclude that this
explains why the scale height found by \citet{gaensler2008} is much
larger than the `canonical' value that was found in the literature up
till then.
Although the selection effect that \citet{savage2009} describe is
real, Fig.~\ref{dmperp_z.fig} shows that it affects only a few of the 45
lines of sight that we used in our Monte Carlo simulation: only the
sightlines towards NGC 5986, NGC 6517, and NGC 6539 have DM$_\perp$
that lie within 1 $\sigma$ (=4.2 cm$^{-3}$pc) of
DM$_{\infty,\perp}$=24.4 cm$^{-3}$pc. Smaller asymptotic values for
DM$_{\infty,\perp}$ are not often encountered in our Monte Carlo
simulation. Furthermore, lines of sight far from the Galactic plane
($|b|$ $>$ 40\degr) consistently reproduce scale heights of about 1.6 $\pm$ 0.3 kpc in our Monte Carlo simulation, which is comparable to the scale
height $h$ = 1.41$^{+0.26}_{-0.21}$ kpc that \citet{savage2009}
themselves find for these latitudes.
We therefore conclude that the scale heights we derive by fitting DM$_\perp$ as a function of $|z|$ instead of the other way round are not strongly affected by the fitting bias that
\cite{savage2009} identified.


\subsection{Parameters for models M2 and M3, discussion of NE2001c}\label{ss-model-m23-details}
As Figs.~\ref{dmperp_z.fig} and \ref{h_n0.fig} showed, an exponential disk model with a single scale height cannot provide the best fit at all latitudes, and the blue and red data points in Fig.~\ref{dmperp_z.fig} that indicate sightlines within 25\degr\ of the Galactic plane show considerable scatter. Furthermore, regions with extended H$\alpha$ emission like the Gum nebula are known to add significant DM to pulsars that lie beyond it. Summarising, the Galactic ISM contains components in addition to the exponential thick disk that we considered so far. In this section and the next we investigate ISM models that model some of these additional components. Since we will follow the same approach to construct models M2 and M3 we will discuss both models here.

Since the exponential thick disk in model M1 has a much larger scale height than the thick disk in TC93 and NE2001 (1.6 kpc vs. 0.9 kpc) we will investigate whether changing the thick disk in TC93 and NE2001 can improve the model performance in predicting DM. To remain internally consistent with the other components in these models we first subtract the DM contributions by all components in TC93 and NE2001 except their thick disks (which we will refer to as the `DM residuals') from the observed DMs. We then fit a new exponential thick disk model to the DM residuals using the procedure described in \S~\ref{ss-model-m1-details}. Contrary to M1, where we used only sightlines at $|b|$ $>$ 32.5\degr, we use all sightlines at $|b|$ $>$ 5\degr\ to determine the scale height of the thick disk in models M2 and M3: we assume here that TC93 and NE2001 model the additional ISM components perfectly, so that subtracting their predicted DMs from the observed DMs (= our definition of the residual DM) leaves only the contribution by the exponential thick disk. This solution is however not perfect. Since the relative magnitudes of the DM contributions by the thin and thick disk (for example) follows from a complex optimisation procedure, changing the contribution of one without changing the contribution of the other can still lead to some internal inconsistencies. 

\begin{figure}
\resizebox{\hsize}{!}{\includegraphics{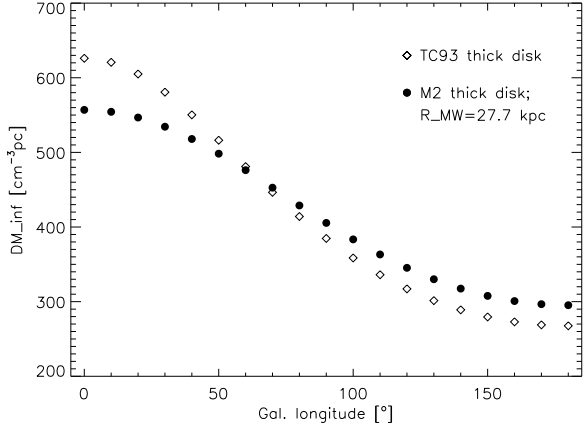}}
\caption{The DM at infinity predicted by the exponential thick disk in model TC93 for $b$=0\degr, and the prediction by the exponential thick disk in model M2 when a radial cut-off at 27.7 kpc is used. We determined this cut-off radius by minimising the vertical separation between the two model curves.
}
\label{dminf_thick_disk_comparison.fig}
\end{figure}

In Fig.~\ref{z_dmsinb_m123.fig} we compare the z vs. DM$_\perp$ predicted by M1--M3 using the best fits to the data points (M1: black data points) and DM residuals (M2, M3: red data points). The black line indicates the best model fit of M1 in all 3 panels, and the red lines show the best fit to the data by model M2 (middle panel) and M3 (right-hand panel). The vertical dotted line indicates the asymptotic value of DM$_{\infty,\perp}$, which is the same for M2 as for M1. The reason for this is that the thin disk, Gum nebula and spiral arms in TC93 do not contribute to the DMs of the 8 sightlines that we used to calculate DM$_{\infty,\perp}$. The local ISM components and local spiral arm that are included in NE2001 do contribute to the DMs of these 8 sightlines, so that correcting for their DM contributions leads to a smaller  DM$_{\infty,\perp}$ = 20.52 $\pm$ 2.72 cm$^{-3}$pc. 

The scale height that we found for the thick disk in M2 is 1587 $\pm$ 309 pc. Since the local ISM components and local spiral arm in NE2001 contribute to the DMs of sightlines at $|b|$ $>$ 40\degr, while GMCM08 assumed that only the exponential thick disk component contributes DM to sightlines this far from the Galactic plane, the exponential thick disk from GMCM08 cannot replace the original thick disk in NE2001. The exponential thick disk that we fitted to the DM residuals for NE2001 has a scale height of 1313 $\pm$ 181 pc, which is in fact \emph{smaller} than the scale heights that we found for M1 and M2, and much smaller than the scale height found by GMCM08. 

To find a physically reasonable radius for the exponential thick disk that we fitted in M2, we require that this thick disk predicts the same DM at infinity as the thick disk in TC93. (We only do this for model M2, since we will show in Sect.~\ref{s-comp} that M2 predicts DM best of all the models that we tested) We first determined the DM predicted at infinity by the thick disk in TC93 for a range of Galactic longitudes, at $b$=0\degr.  We then minimised the mean square difference between the DM at infinity predicted by the thick disk components in TC93 and M2 for a range of trial radii. By fitting a parabola to this minimum we found that a radius of 27.7 kpc gives the overall best match between the DM predicted by the thick disks in TC93 and M2. The difference between the model predictions at infinity increases at very small and very large Galactic longitudes, see Fig.~\ref{dminf_thick_disk_comparison.fig}. The reason for this difference is that while the exponential disk in model M2 has a uniform mid-plane electron density, the thick disk in TC93 shows a radial dependence of the electron density that is proportional to sech$^2$(radius).

\begin{figure}
\resizebox{\hsize}{!}{\includegraphics{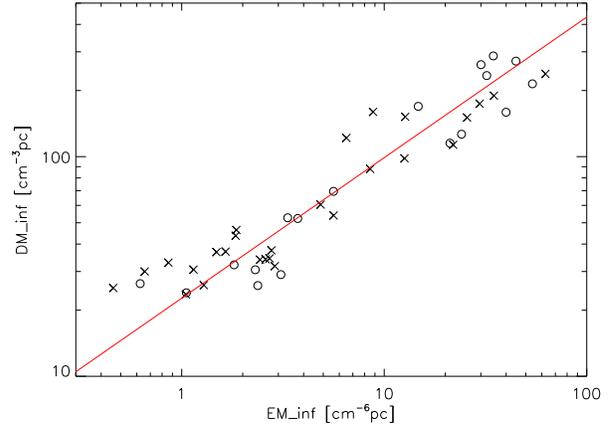}}
\caption{DM$_\infty$ vs. EM$_\infty$ for the lines of sight towards the selected pulsars from Table \ref{psr.table} (crosses) and globular clusters from Table \ref{psr_gc.table} (circles). The measured DMs were extrapolated to infinity using model M2 and include the remainder after fitting the thick disk in M2 to the residual DMs. The EM$_\infty$ were calculated using the E(B-V) from \citet{schlegel1998}. The best bisector fit DM$_\infty=22.67\ \mathrm{EM}_\infty^{0.64}$ is indicated by a red line. 
}
\label{dminf_eminf.fig}
\end{figure}

\subsection{Parameters for model M4}\label{ss-model-m4-details}
In Fig.~\ref{dminf_eminf.fig} we plot the DM at infinity, DM$_\infty$, against the EM$_\infty$ that we calculated from observed H$\alpha$ intensities (\S~\ref{ss-calculating-em}). The DM$_\infty$ were calculated as the sum of the prediction by model M2 at infinity plus the difference between the residual DMs and the model prediction by the thick disk in M2 for these residual DMs. 
If model M4 can accurately predict this difference then it can outperform M2. (We used model M2 to predict DM at infinity since we will show in Sect.~\ref{s-comp} that M2 predicts DM most accurately of all the models that we tested) Since we do not know a priori which ISM component is responsible for the the difference between the residual DM and the DM predicted by the thick disk in M2, we cannot predict how this term changes when we extrapolate it to infinity.

Interestingly, DM$_\infty$ and EM$_\infty$ show a correlation over 2 orders of magnitude, for both
isolated pulsars and pulsars in globular clusters. The existence of this correlation is not a new result,
since it was already described by \citet{berkhuijsen2006}. With an OLS
bisector method \citep[see the instructive review in][]{feigelson1992}
we found that the best fitting power law is described by

\begin{eqnarray}
\mathrm{DM}_\infty\ =\ (22.67\ \pm\ 2.20)\ \mathrm{EM}_\infty^{0.64\ \pm\ 0.03}
\end{eqnarray}

\noindent With our model M4 we investigate whether we can use this correlation
to predict DM by using the observed EM as a proxy.

We established
the 1-$\sigma$ errors of the coefficients of the
DM$_\infty$-EM$_\infty$ fit by running a Monte Carlo simulation of the
distribution of the asymptotic values of DM$_{\infty,\perp}$. For each
realisation we determined the scale height for the thick disk in M2 (\S~\ref{ss-model-m23-details}), which we then used to calculate
DM$_\infty$ from the measured DMs. By using a residual resampling
bootstrap method we then determined the errors in the coefficients of
the DM$_\infty$ -- EM$_\infty$ fit, and we used the ensemble of all
Monte Carlo realisations to establish the overall errors in these
coefficients (viz. the errors that include both the scatter in the
asymptotic DM$_{\infty,\perp}$, and the scatter in the fitted
DM$_\infty$ -- EM$_\infty$ relation that is due to structure in the
ISM). We checked that these errors did not change significantly when
we changed the number of Monte Carlo realisations or the number of
resamplings used in the bootstrap method.

\section{Comparison between model predictions and the data}\label{s-comp}

\begin{figure*}
\centering
\resizebox{0.84\hsize}{!}{\includegraphics{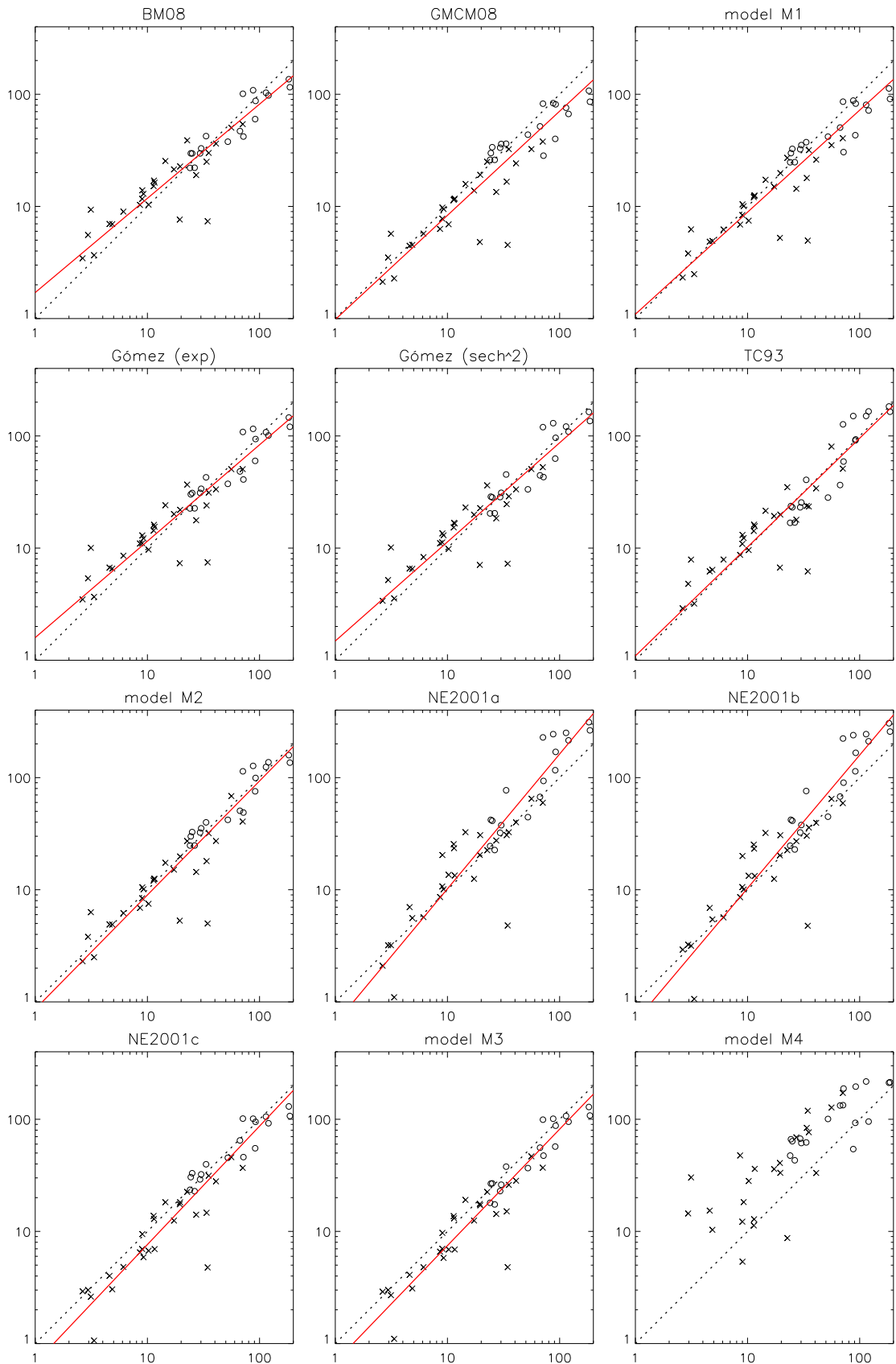}}
\caption{Comparison between the modelled and measured DMs (vertical and horizontal axes) for the 12 models from Table \ref{DMmodels.table}. Isolated pulsars are indicated by crosses, pulsars in globular clusters as open circles. The red line shows the OLS bisector fit to the data points. For model M4 the least-squares fitting algorithm did not converge (which is probably related to the very large scatter of the data points) so that no OLS bisector line was drawn in that case.
}
\label{DMmodel_vs_DMobs.fig}
\end{figure*}

The predictions for DM made by the various models will not be perfect;
instead they will scatter around the DMmodel = DMobserved line
(Fig. \ref{DMmodel_vs_DMobs.fig}; we will abbreviate `DMobserved' to
`DMobs' from here on). When analysing the scatter that the different
models produce, one considers models with smaller fractional
deviations to be better than models with small absolute deviations; we
therefore plotted Fig. \ref{DMmodel_vs_DMobs.fig} with log axes. This
figure also shows the ordinary least squares bisector fit of
log(DMmodel) as a function of log(DMobs) as a red line.
All models apart from TC93 clearly show a systematic deviation
between the red line and the DMmodel = DMobs line (the dotted line).

In this section we will investigate which model predicts the observed
DMs most accurately over the entire range of observed DMs. Accuracy entails both the absence of systematic deviations and a small
scatter around the DMmodel = DMobs line. Although models may be very
accurate over a small range in DM, where the red line intersects the
dotted line, we will consider the predictive quality of the models
over the full distance range, i.e., from nearby pulsars to pulsars in distant
globular clusters. 

\begin{figure*}
\resizebox{0.84\hsize}{!}{\includegraphics{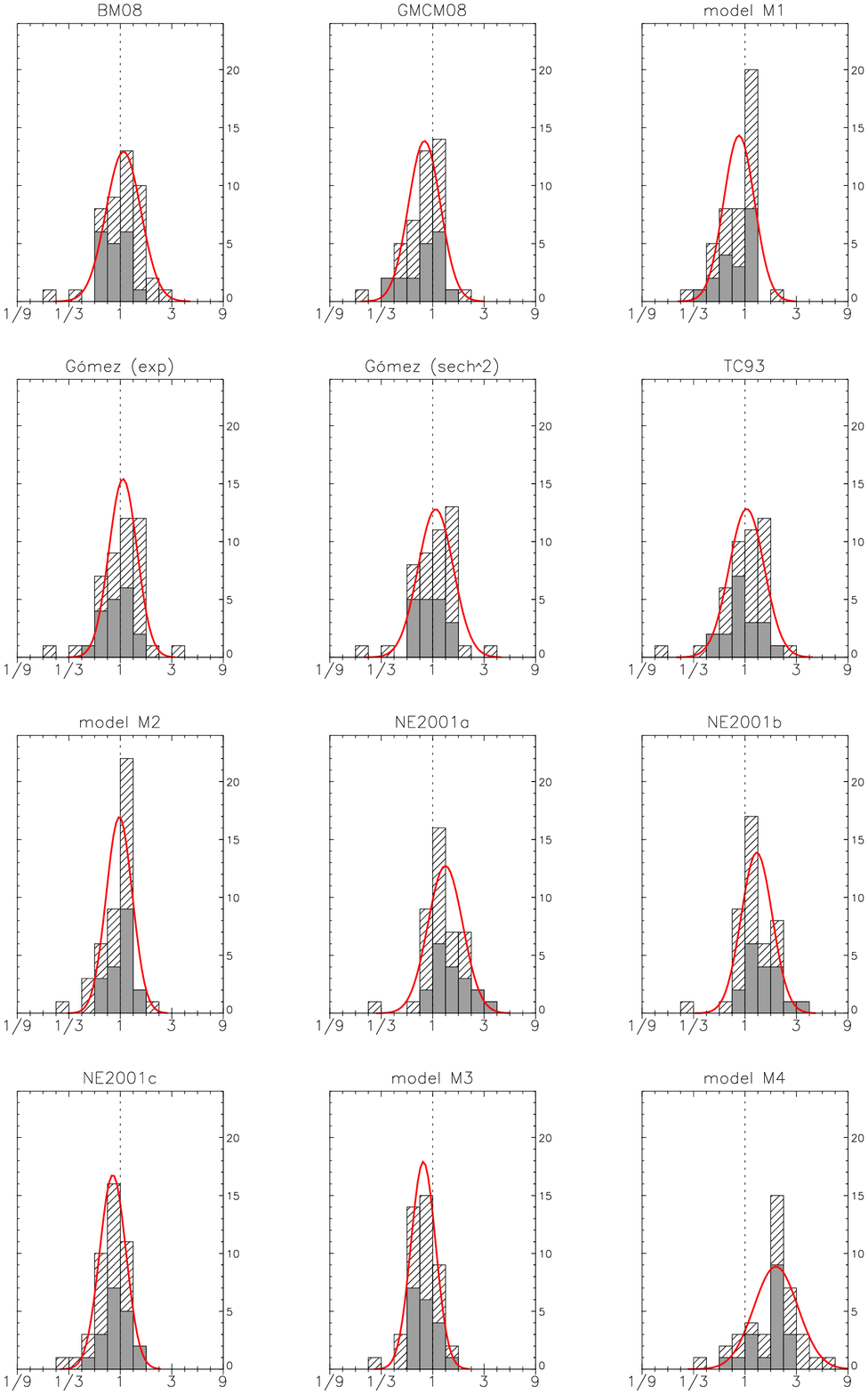}}
\caption{
Histograms of the ratios between the modelled and observed DMs for each of the 12 models from Table \ref{DMmodels.table}.   The hatched and shaded bins indicate data from isolated pulsars and from pulsars in globular clusters respectively. The red curve shows the normal distribution whose mean and width were determined using robust statistics, the amplitude of the curve then follows from a least-squares minimisation. 
}
\label{ratio_distr.fig}
\end{figure*}

One can define different ways for quantifying a (fractional) deviation
between DMmodel and DMobs. An obvious choice is DMmodel/DMobs, and
other possibilities are 
(DMmodel-DMobs)/DMobs, and (DMmodel-DMobs)/($\sqrt{2}$ DMobs), which
respectively quantify the vertical distance and the perpendicular
distance between a point and the line DMmodel=DMobs.  As the latter two
are both proportional to DMmodel/DMobs, the distribution that shows
the smallest scatter in DMmodel/DMobs also shows the smallest scatter
for the other 2 choices for quantifying deviations: the way one
calculates the deviation between DMmodel and DMobs is not critical
when defining what constitutes the best model.


In Fig. \ref{ratio_distr.fig} we show histograms of the distribution
of DMmodel/DMobs for the 45 lines of sight that were selected in \S~\ref{ss-los-selection}, and we indicate data towards solitary pulsars with
hatched bins, and data from pulsars in globular clusters with shaded
bins. We used the subset of 45 lines of sight so that we can include model M4 in our comparison. Later in this section we will present results for all 68 available lines of sight.
To emphasize the importance of fractional deviations between
DMmodel and DMobs, we plotted the ratio DMmodel/DMobs using a
logarithmic x-axis. Systematic deviations between the modelled and
measured DMs show up in this figure as different distributions for the
filled and hatched bins, since the pulsars in globular clusters have
much larger DMobs than the solitary pulsars. The panels in this figure
also show the normal distributions that were fitted to these data sets
(red lines).
The mean and spread of the fitted Gaussian curve were calculated from
the distribution of DMmodel/DMobs of the 45 lines of sight; the
amplitude of the fitted curve guarantees the correct normalization of
the distribution.

\begin{figure}
\resizebox{\hsize}{!}{\includegraphics{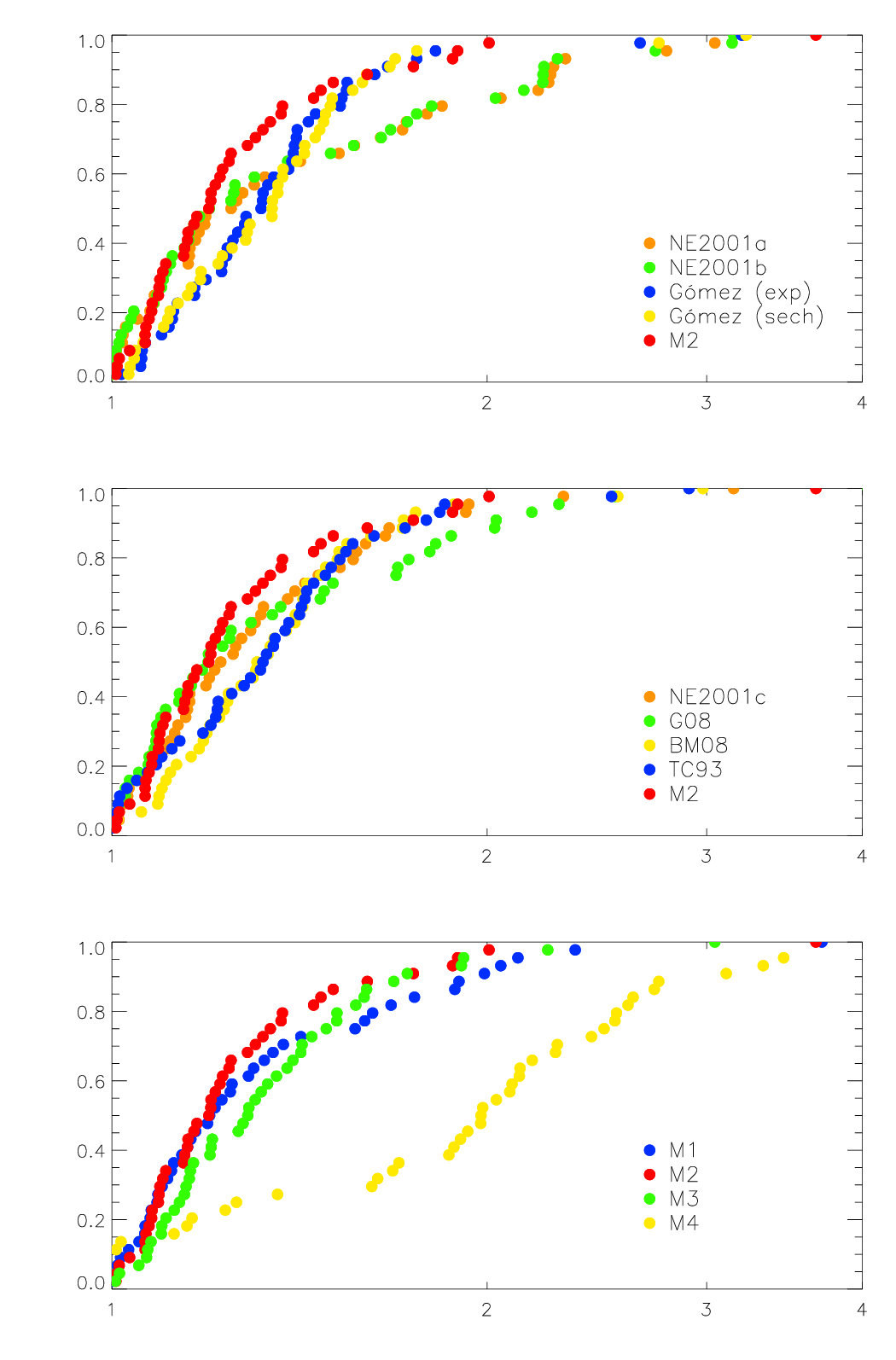}}
\caption{Cumulative distribution of the number of lines of sight where the modelled DM lies within a factor of N of the observed DM. We only plot N for the 45 lines of sight selected in \S~\ref{ss-los-selection} so that model M4 can be included, and we used 3 panels to avoid cluttering the plot. 
}
\label{cumulative.fig}
\end{figure}

\begin{figure}
\resizebox{\hsize}{!}{\includegraphics{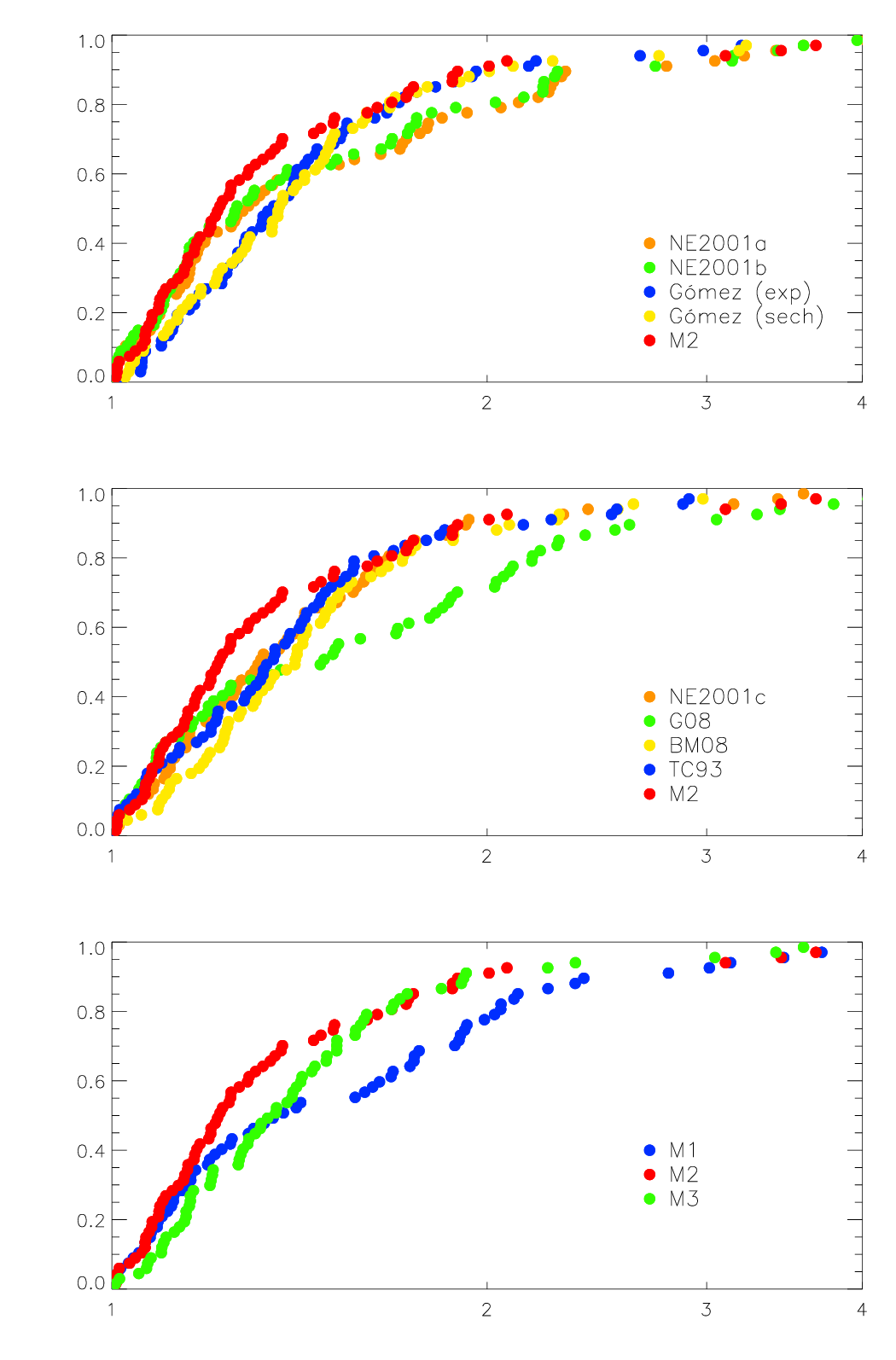}}
\caption{The equivalent of Fig.~\ref{cumulative.fig} for all 68 available lines of sight from Tables \ref{psr.table} and \ref{psr_gc.table}. 
}
\label{cumulative_los68.fig}
\end{figure}

As is clear from comparing the histograms to the fitted normal
distributions, the latter are not a very good description of the data
points.  This means that the quality of the models can not be
described in terms of a mean and spread of the fitted Gaussian. Also,
the concept of statistical outliers is ill-defined since sigma is not
a good measure of the width of the distribution. Therefore, to compare
the quality of the different models, we count for each model how many
lines of sight have a DMmodel that deviates by less than a factor of N
from DMobs (so N = DMmodel/DMobs when that ratio is larger than 1, otherwise N = DMobs/DMmodel)
for different N -- the
result is the cumulative distributions of Figs.~\ref{cumulative.fig} and \ref{cumulative_los68.fig},
where N is plotted along the x-axis. The best model should for each source have the smallest N compared to
other models. This translates into a cumulative distribution in
Figs.~\ref{cumulative.fig} and  \ref{cumulative_los68.fig} that rises more steeply than the other
distributions (`which lies above the other distributions' in these figures). This way of counting
automatically incorporates the systematic offsets between modelled and
observed DMs that are seen in Fig. \ref{DMmodel_vs_DMobs.fig}. Increasing the thick disk in TC93 has produced a model that predicts DM more accurately than the other models, both for the subset of 45 `good' sightlines, and for the full sample of 68 sightlines.  



In Table \ref{DMmodels.table} we list for each of the models the
values of N within which 50\%, 75\% and 90\% of the lines of sight
lie. This table and Figs.~\ref{cumulative.fig} and \ref{cumulative_los68.fig} show that all
models can predict DM to within a factor of 1.5--2 for 3/4 lines of sight. The DMs of sightlines that lie further than 5\degr\ from the Galactic plane, and avoid local features in the H$\alpha$ maps, can be predicted even more accurately, to within 40--70\% for 3/4 sightlines. M4 fails to be as accurate as the other models. The scatter around the DM$_\infty$(EM$_\infty$) relation probably predicts a DM$_\infty$ that, when corrected for the finite distance of the pulsar, produces a DM which deviates strongly from the observed DM. Model M4 might be
improved upon, either by finding a second parameter that tightens the
DM$_\infty$--EM$_\infty$ fit, or by considering a different proxy.

The difference in predictive quality
between the different models becomes more apparent when considering
the quality with which the DMs of 75\% and 90\% of the lines of sight
are predicted. Excluding M4, the most accurate models predict DM to within a factor of 2 of the observed DM, while the poorest performing models predict DM within a factor of 3 for 9/10 of sightlines. The DMs of the 23 sightlines that lie within 5\degr\ of the Galactic plane or that lie towards structures in H$\alpha$ intensity  are not by definition difficult to predict. It is somewhat surprising to find that the simpler models perform almost as well as the more complex models that are based on TC93 or NE2001. This is probably related to how the pulsars that we use to test the different models are distributed through the Milky Way. 

%

\section{Sightlines with anomalous N}\label{s-anomalousn}

The cumulative distributions of N also lend themselves for defining the number
of `outliers' for which the modelled DMs deviate strongly from the
measured DMs, for example by looking up the fraction of lines of sight
with N $>$ 2. In Table \ref{Outliers.table} we list for each of the models 
which lines of sight have N that lie above the 90th percentile, supplemented with the lines of sight that have N $>$ 2. 
Although in general different models have difficulty in predicting the DMs of different sightlines, most models cannot accurately predict the DM of J0358+5413, J0630-2834,  J0659+1414, J0826+2637, J0835-4510, J1643-1244, and J1909-3744. Only in the case of J1909-3744 is the modelled DM larger than the observed DM, for all models. The DM that were calculated for the other sightlines are smaller than the observed DM.

When the high-resolution H$\alpha$ maps from SHASSA or
VTSS indicate an \hii\ region in the direction of one of our lines of
sight, we can use that line of sight to put a upper or lower limit on
the distance to the \hii\ region. If the observed DM is larger than the predicted DM,
the real ISM contains more structure than is included in the model,
which puts the pulsar or globular cluster beyond the \hii\
region. Similarly, when the sightline does not show an anomalous N,
the pulsar or globular cluster must lie in front of the \hii\ region. This way the distances that were determined for J0358+5413, J0659+1414 and J1643-1244 give upper limits to the \hii\ regions that must lie in front of them. The proximity of J0358+5413 to the Galactic plane, combined with a sightline of over 1 kpc, could mean that more than one \hii\ region contributes to its DM. Although J0835-4510 is seen towards the Gum and Vela nebulae, its distance places it in front of these nebulae (see also\ \citealt{mitra2001}). It turns out that models which are based on NE2001 \emph{under}predict the DM of this pulsar, while the other models overpredict the DM.

We analysed the high-resolution
SHASSA and VTSS maps of H$\alpha$ intensity towards J0630-2834, but we did not find strong H$\alpha$ emission in the
vicinity of these lines of sight. J0826+2637 is only covered by the
low-resolution WHAM data.
%
There is however diffuse H$\alpha$ emission over a wider region
surrounding the lines of sight towards J0630-2834 and J0826+2637, but to identify such regions would
require more advanced selection tools.

\begin{table*}
\centering
\caption{Overview of sightlines that lie above the 90th percentile of the cumulative distribution of N (values of N indicated in italics), supplemented with sightlines for which N $>$ 2 (i.e. DM$\_$model $>$ 2$\times$DM$\_$obs, or  DM$\_$model $<$ 1/2$\times$DM$\_$obs). $\underline{\mathrm{Name}}$: line of sight was not part of the 45 lines of sight that were selected in \S~\ref{ss-los-selection}. Since we are investigating the N of all 68 lines of sight we left out the results for model M4, which can only predict DMs well for the subset of 45 lines that were selected in the aforementioned section. In the final column we list the Galactic latitudes of the lines of sight, rounded off to the nearest integer.
}
\begin{tabular}{l|ccccccccccc|r}
\hline
Sightline                             &    BM08          & GMCM08       &          M1         &    \multicolumn{2}{c}{GBC}     & TC93            &          M2          &           \multicolumn{3}{c}{NE2001}                    &         M3       & \multicolumn{1}{c}{Gal.} \\
                                             &                         &                         &                        &   exp(-z)        & sech$^2$(z)    &                       &                          &             a         &          b              &         c              &                      & \multicolumn{1}{c}{latitude} \\ \hline
J0034-0721                       &      --                &     --                 &     --                 &     --                &     --                   &     --                &    --                   &   2.24              &    2.22              &     --                 &       --            & -70\degr \\
\underline{J0139+5814} &      --                &   2.10              &     --                 &     --                &     --                   &     --                &    --                   &    --                  &     --                  &     --                 &       --            & -4\degr \\
\underline{J0358+5413} & \emph{2.28}  & \emph{3.79}  & \emph{3.46} &    2.16            &     2.10              & \emph{2.87} &  \emph{3.44} &    --                  &     --                  &     --                 &       --            &  1\degr \\
J0454+5543                      &      --                &     --                 &     --                 &     --                &     --                   &     --                &    --                   &   2.25              &    2.22              &     --                 &       --            &   8\degr \\
\underline{J0538+2817} &      --                &    2.08             &     --                 &     --                &     --                   &     --                &    --                   &     --                 &     --                  &     --                 &       --            &  -2\degr \\
J0630-2834                       &  \emph{4.67} & \emph{7.60} & \emph{6.95}  & \emph{4.63} & \emph{4.75}    & \emph{5.56} & \emph{6.89}  & \emph{7.18}  & \emph{7.22}  & \emph{7.22}  & \emph{7.18} & -17\degr \\
\underline{J0659+1414} &     2.08            & \emph{3.43} & \emph{3.14}  &     --                &     --                   & \emph{2.54} & \emph{3.11}   & \emph{3.41}  & \emph{3.42}  & \emph{3.42}  &  \emph{3.41} &  8\degr \\
\underline{J0737-3039A/B} &      --           &   3.06             &   2.80              &     --                &     --                   &  --                   &    --                   &   2.12              &     --                  &     --                 &       --            &  -5\degr \\
J0826+2637                      & \emph{2.55}  & \emph{4.04} & \emph{3.71}  & \emph{2.65} & \emph{2.75}   & \emph{2.90} & \emph{3.67}   &    --                  &     --                  &     --                 &       --            &  32\degr \\
\underline{J0835-4510}  & \emph{10.34} & \emph{17.17} & \emph{15.66} & \emph{9.11} & \emph{8.59} & \emph{12.36} & \emph{15.11} & \emph{3.59} & \emph{3.59} & \emph{3.59}  & \emph{3.59} &  -3\degr \\
J0922+0638                      &      --                &    2.03            &     --                  &     --                &     --                   &     --                &    --                   &     --                 &     --                  &     --                 &       --            &  36\degr \\
J1012+5307                      &      --                &     --                &     --                  &     --                &     --                   &     --                &    --                   &    2.26             &   2.22               &     --                 &       --            &  51\degr \\
\underline{J1643-1244}  & \emph{6.04} & \emph{9.61} & \emph{8.83}   & \emph{6.12} & \emph{6.31}   & \emph{6.86} & \emph{8.79}   &  \emph{16.00} & \emph{15.87} & \emph{22.92} & \emph{22.29} &  21\degr \\
J1738+0333                      &      --                &    2.03            &     --                  &     --                &     --                  &     --                 &    --                   &     --                 &     --                  & \emph{2.30}  & \emph{2.24} &  18\degr \\
J1744-1134                       & \emph{2.98}  &     --                &      --                 & \emph{3.20} & \emph{3.23}   & \emph{2.52} & 2.01                 &     --                 &     --                  &     --                 &        --           &   9\degr \\
\underline{J1857+0943} &      --                &     --                &      --                 &     --                &    2.01               &     --                &    --                   &     --                 &     --                  &     --                 &        --           &  3\degr \\
\underline{J1909-3744}  &     2.28            &     --                &       --                & \emph{2.19} & \emph{2.26}   &    2.25            &    --                   &  \emph{4.02} & \emph{3.96}  &     --                  &       --           &  -20\degr \\
\underline{J1932+1059} &  \emph{2.62} &     --                &      --                 & \emph{2.98} & \emph{3.19}   &   2.14             &    --                   &     --                 &     --                  &     --                 &       --            &  -4\degr \\  
J2048-1616                       &      --                &     --                &       --                &     --                &     --                   &     --                &    --                   &   2.05              &    2.03              &     --                 &       --            & -33\degr \\
J2144-3933                       &      --                &     --                &      --                 &     --                &     --                   &     --                &     --                  & \emph{3.05}  & \emph{3.15}   & \emph{3.15}  &  \emph{3.05} & -49\degr \\
\hline
\underline{NGC 6121}     &      --                &    2.40           &   2.24               &     --                &     --                   &     --                &    --                   &     --                 &     --                  &     --                  &       --            &  16\degr \\
NGC 6266                          &      --                &     --               &     --                   &     --                &     --                   &     --                &    --                   &    2.20            &    2.14              &     --                  &       --            &   7\degr \\
NGC 6342                          &      --                &     --               &     --                   &     --                &     --                   &     --                &    --                   & \emph{3.22} & \emph{3.14}   &     --                  &       --            &  10\degr \\  
NGC 6397                          &      --                &    2.53           &    2.35              &     --                &     --                   &     --                &    --                   &     --                 &     --                  &     --                  &       --            & -12\degr \\
\underline{NGC 6440}     &      --                &    2.18           &    2.03              &     --                &     --                   &     --                &    --                   &     --                 &     --                  &     --                  &       --            &   4\degr \\
\underline{NGC 6522}     &      --                &    2.06           &     --                  &     --                &     --                   &     --                &    --                   &     --                 &     --                  &     --                  &       --            &  -4\degr \\
NGC 6539                          &      --                &    2.17           &    2.05              &     --                &     --                   &     --                &    --                   &     --                 &     --                  &     --                  &       --            &   7\degr \\
\underline{NGC 6544}     &     2.04            & \emph{3.29} & \emph{3.01}  &     --                &     --                   &     --                &    --                   &     --                 &     --                  &     --                  &       --            &  -2\degr \\
NGC 6624                          &      --                &     --               &     --                   &     --                &     --                   &     --                &    --                   &   2.79             &    2.73              &     --                  &       --            &  -8\degr \\  
NGC 6656                          &      --                &   2.28            &    2.12              &     --                &     --                   &     --                &    --                   &     --                 &     --                  &     --                  &       --            &  -8\degr \\
\underline{NGC 6749}     &      --                &     --               &     --                   &     --                &     --                   &     --                &    --                   &   2.29             &    2.27              &     --                  &       --            &  -2\degr \\
NGC 6752                          &      --                &     --               &     --                   &     --                &     --                   &     --                &    --                   &   2.31             &    2.28              &     --                  &       --            & -26\degr \\
\underline{NGC 6760}     &      --                &    2.21           &    2.05              &     --                &     --                   &     --                 &    --                  &     --                 &     --                  &     --                  &       --            &  -4\degr \\
\underline{NGC 6838}     &      --                &    2.28           &    2.10              &     --                &     --                   &     --                 & \emph{2.07}  &     --                 &     --                  & \emph{2.41}   & \emph{2.35} & -5\degr \\
\underline{Terzan 5}        &      --                &    2.60           &    2.39              &     --                &     --                   &     --                 &    --                   &     --                &     --                   &     --                  &       --           &   2\degr \\  
\hline
\end{tabular}
\label{Outliers.table}
\end{table*}

We did not exclude these lines of sight from our analysis since we want to keep our selection criteria simple. In practice, since these lines of sight have the highest N in most models, they affect the cumulative distributions of these models in the same way, and only above the 90th percentile.

\section{Conclusions}\label{s-conclusions}
In this paper we have analysed how well twelve different models of the Galactic free electron density $n_{\mathrm{e}}$ predict the measured DMs of pulsars at well-known distances. These distances were derived from either parallax measurements, or from association of the pulsar with a globular cluster. Eight of these models have already been published in the literature. We determined a scale height of 1.6 kpc for our exponential thick disk model M1, where we used only sightlines with $|b|$ $>$ 32.5\degr\ to avoid the more complex structure at smaller Galactic latitudes. In models M2 and M3 we replaced the thick disk from the models by \citet{taylor1993} and \citet{cordes2002} by exponential thick disks. First we had to correctly account for the DM contributions by the other components in these models, which we did by subtracting these DMs from the observed DMs before fitting a new scale height for the thick disk. The thick disks in M2 and M3 have scale heights of 1.6 and 1.3 kpc, resp. When we limit the radial extent of the thick disk in model M2 to 27.7 kpc, it produces for $|b|$=0\degr\ the same DM at infinity as the original thick disk from the model by \citet{taylor1993}. Although this radius on average reproduces the DM of the original thick disk, the DM difference between the old and new thick disk varies somewhat as a function of Galactic longitude. With model M4 we explore using the emission measure as a proxy to predict the observed scatter in DM that the other models cannot explain. 

A few of the components in the model by \citet{cordes2002} contribute also to the DMs of sightlines that lie further than 40\degr\ from the Galactic plane. Since \citet{gaensler2008} only fitted an exponential thick disk, incorporating their model for the thick disk in the model by \citet{cordes2002} leads to internal inconsistencies. 

We developed a novel approach for
comparing the predictive quality of the different models, by first
calculating for each sightline the factor N by which the modelled DM
deviates from the observed DM (so N = modelled DM/observed DM when that ratio is larger than 1, otherwise N = observed DM/modelled DM), and then comparing the cumulative distributions of N between the
models. Most models can predict DM to within a factor of 1.5-2 of the measured value for 3/4  lines of sight, and the model by \citet{taylor1993} that we updated with a more extended thick disk consistently performs better than the other models that we tested. Our fourth model predicts DM least accurately. Surprisingly, the simpler models that we tested perform almost as well as the more complex models. This is probably related to the sparse spatial distribution of the sightlines that we used. Finally, by comparing which pulsar DMs are difficult to predict by most models we identified pulsars that lie beyond \hii\ regions. The accurate pulsar distances then place upper limits on the distances to these \hii\ regions. 

The analysis we present here will benefit from having a larger sample of sightlines with known distances and dispersion measures, which can come from either parallax measurements of additional pulsars, or from discoveries of more globular clusters with pulsars in them. A different approach in the modelling philosophy, for example by
including H$\alpha$ data or a different proxy, warrants further
investigation, as it can help explain the detailed structure in DM.


\section*{Acknowledgements}
First, I would like to thank Peter Katgert (Sterrewacht Leiden) for the time and effort he put into our discussions on this topic. I would also like to thank Paulo Freire (MPIfR Bonn), Ryan Lynch (UVA, now at McGill University), Simon Johnston (ATNF/CASS), Kejia Lee (MPIfR Bonn), Joris Verbiest (MPIfR Bonn), and Aristeidis Noutsos (MPIfR Bonn) for their help during different phases of this project, and Simon Johnston (ATNF/CASS) and Michael Kramer (MPIfR Bonn) for proof-reading the manuscript. I would like to thank the staff at Leiden Observatory, and the
Leidsch Kerkhoven-Bosscha Fonds, for their generous support of this
project. The Wisconsin
H-Alpha Mapper (WHAM), the Southern H-Alpha Sky Survey Atlas (SHASSA)
and the Virginia Tech Spectral-Line Survey (VTSS) are all funded by
the National Science Foundation. To obtain the extinction data by
Rowles \& Froebrich we used their extinction map query page, which is
hosted by the Centre for Astrophysics and Planetary Science at the
University of Kent. During this project I moved from ATNF/CASS to the MPIfR, and I worked on this project at both institutes. The order of my affiliations is purely chronological.


\bibliography{ne6e} 

\end{document}